\documentclass[aps,reprint,nofootinbib,nobibnotes,notitlepage,superscriptaddress,onecolumn,prd,
 amsmath,amssymb
]{revtex4-2}

\usepackage[caption=false]{subfig}
\usepackage{braket}
\usepackage{float}
\usepackage{lipsum}
\usepackage{graphicx}
\usepackage{dcolumn}
\usepackage{bm}
\usepackage{natbib}
\usepackage{hyperref}
\hypersetup{
	colorlinks = true,
    linkcolor = Red,
    urlcolor  = Red,
    citecolor = Red
}
\usepackage{subfig}

\usepackage{microtype}

\usepackage{verbatim}
\usepackage[amssymb]{SIunits}
\usepackage{tabularx}
\usepackage[dvipsnames]{xcolor}
\usepackage{wasysym}
\usepackage[export]{adjustbox}
\usepackage{multirow}
\usepackage{tikz}

\def\be{\begin{equation}}
\def\ee{\end{equation}}
\usepackage{enumitem}

\usepackage{color}
\definecolor{darkgreen}{RGB}{0,120,0}

\definecolor{darkgreen}{RGB}{0,120,0}

\newcommand{\delD}[1]{(2\pi)^3\delta_\mathrm{D}\left({#1}\right)}

\newcommand{\av}[1]{\left\langle{#1}\right\rangle} 

\newcommand{\vk}{\vec k}

\def\beq{\begin{eqnarray}}
\def\eeq{\end{eqnarray}}
\let\vec\mathbf

\usepackage{empheq}

\usepackage{booktabs}

\begin{document}

\title{Dissecting the Scalar Cosmological Collider with the Cosmic Microwave Background}


\author{Oliver~H.\,E.~Philcox}
\email{ohep2@cantab.ac.uk}
\affiliation{Leinweber Institute for Theoretical Physics at Stanford, 382 Via Pueblo, Stanford, CA 94305, USA}
\affiliation{Kavli Institute for Particle Astrophysics and Cosmology, 382 Via Pueblo, Stanford, CA 94305, USA}

\begin{abstract} 
\noindent 
The cosmological collider program probes ultra-high-energy physics by searching for subtle oscillatory signatures induced by inflationary particle exchange. Due to the non-linear symmetries usually assumed in inflation, these signals do not appear in isolation; moreover, their amplitudes are bounded by perturbativity and unitarity. To comprehensively probe cosmological collider physics, we must jointly analyze the full multi-field inflationary Lagrangian: in this work, we conduct such a study, probing single, double, and triple scalar field exchange and self-interactions across a wide range of masses (in both the complementary and principal series), mixings, and sound-speeds. Using modern theoretical tools (including the cosmological bootstrap and the cosmological flow), we construct a vast library of tree-level primordial bispectra: combining these with recent measurements of the inflationary shape function from \textit{Planck} and modern sampling techniques, we perform Bayesian exploration of the eight-parameter multi-field likelihood. Most previous studies of cosmological collider physics assume weak mixing, such that the coupling between the scalar field and the Goldstone mode can be treated perturbatively: in our companion study, we demonstrate that this assumption is incompatible with current datasets except at the smallest masses. By solving for the inflationary bispectra numerically using \textsc{CosmoFlow}, we perform the first analysis of the strongly-mixed collider, demonstrating that the \textit{Planck} constraints are much tighter than the theoretical bounds, though the oscillatory contributions are heavily suppressed. Across the full multi-field landscape, we find no evidence for new physics with a maximal $\chi^2$ improvement of $5.3$.
\end{abstract}

\maketitle

\section{Introduction}\label{sec: intro}
\noindent During inflation, many things can happen at once. Whilst the simplest models feature just a single field slowly rolling down its potential \citep[e.g.,][]{Guth:1980zm,Starobinsky:1980te,Linde:1981mu,1982PhRvL..48.1220A,Mukhanov:1981xt,Starobinsky:1982ee}, the literature abounds with examples of non-standard behavior, involving jumps in the potential \citep{Flauger:2009ab,Flauger:2010ja,Adshead:2011jq,Achucarro:2012fd}, interactions between inflationary fields \citep{Langlois:2008qf,Senatore:2010wk,Byrnes:2010em,Chen:2009zp}, particle production \citep{Flauger:2016idt,Sou:2021juh,Kim:2021ida,Barnaby:2009dd,Lee:2016vti,Arkani-Hamed:2015bza}, dissipation \citep{Berera:1995ie,LopezNacir:2011kk,Salcedo:2024smn}, and beyond. Many of these phenomena leave traces in the distribution of fluctuations at the end of inflation, in the form of transiently-broken scale invariance \citep{Chen:2010xka,Chen:2015lza}, non-Gaussian correlations \citep{Maldacena:2002vr,Bartolo:2004if,Komatsu:2010hc,Liguori:2010hx,Chen:2010xka,Meerburg:2019qqi,Achucarro:2022qrl,Arkani-Hamed:2015bza,Chen:2009zp,Lee:2016vti}, and isocurvature modes \citep{Lyth:2001nq,Dvali:2003em,Bartolo:2003jx,Sasaki:2006kq,Suyama:2007bg}, which has motivated a wide range of searches using cosmic microwave background (CMB) \citep[e.g.,][]{Creminelli:2005hu,Komatsu:2003iq,Senatore:2009gt,Philcox:2025wts,Philcox:2024jpd,El-Haj:2025zbe,Fergusson:2010gn,Philcox:2026njr,2014A&A...571A..24P,Planck:2015zfm,Planck:2019kim,Sohn:2024xzd,Jung:2025nss,Salcedo:2026sdn,Suman:2025vuf,Suman:2025tpv,Kumar:2026ogn,Kumar:2026dih,Philcox:2025bbo,Philcox:2026bfa} and large-scale structure (LSS) \citep[e.g.,][]{Cabass:2022wjy,Cabass:2022ymb,DAmico:2022gki,Green:2023uyz,Cabass:2024wob,Chudaykin:2025vdh,Green:2026yev,Chaussidon:2024qni} datasets. To date, there have been no robust detections.

How can we systematically study the inflationary paradigm? A powerful approach is to invoke certain symmetries (e.g., dilatation invariance) and build a bottom-up model for the primordial Universe using effective field theories (EFTs) \citep{Cheung:2007st,Weinberg:2008hq,Senatore:2010wk}. These provide a powerful tool for categorizing inflationary interactions through the construction of an effective Lagrangian for the low-energy degrees-of-freedom at a given order in fields and their derivatives; this is particularly useful since each of the interaction vertices can be mapped to a specific correlation function at the end of inflation. This approach has been the subject of a number of recent works, which bound the `cosmological collider' couplings of massive (and possibly spinning) fields to the inflaton \citep{Chen:2009zp,Chen:2009we,Baumann:2011nk,Assassi:2012zq,Noumi:2012vr,Chen:2012ge,Pi:2012gf,Green:2013rd,Sefusatti:2012ye,Gong:2013sma,Arkani-Hamed:2015bza,Chen:2016hrz,Flauger:2016idt,Arkani-Hamed:2018kmz,Kumar:2018jxz,Hook:2019vcn,Hook:2019zxa,Wang:2019gbi,Bodas:2020yho,Lu:2021gso,Reece:2022soh,Craig:2024qgy,McCulloch:2024hiz,Jiang:2025mlm,Pimentel:2025rds,Kumar:2025anx,Colas:2025ind,Green:2026yev,Ferreira:2026tyj,Aoki:2026olh,You:2026xoq,Arundine:2026myr,Aoki:2026vbc} by searching for the induced non-Gaussian correlation functions in the CMB temperature and polarization anisotropies \citep{Sohn:2024xzd,Suman:2025tpv,Suman:2025vuf,Philcox:2026bfa,Philcox:2026njr,Kumar:2026dih,Kumar:2026ogn,Salcedo:2026sdn,Cassem:2026ygh}.

Conventionally, primordial non-Gaussianity studies proceed by splitting the inflationary signal into a normalized template, $S_\zeta$ (encoding the type of interaction) and a scaling amplitude, $f_{\rm NL}$ (encoding the Lagrangian coupling) \citep{Komatsu:2001rj,Babich:2004gb,Senatore:2009gt}. Typically, one chooses an inflationary model, computes $S_\zeta$ using analytic or numerical techniques, combines with observational data to constrain $f_{\rm NL}$, then repeats \textit{ad nauseum}. A limitation of this approach is that the inferred $f_{\rm NL}$ may not be consistent with theoretical bounds; as demonstrated in \citep{cosmoflow2} (and reaffirmed below) observable signatures from the single exchange of a massive scalar require non-perturbatively large couplings across most of the parameter space. Furthermore, template-based analyses usually assume that each signal is independent and can be separately searched for in data. This is not necessarily true: non-linearly realized symmetries induce correlations between different inflationary vertices and, on naturalness grounds, one expects that multiple operators should have non-negligible amplitudes. A complete treatment requires overcoming both limitations: we must consistently and simultaneously constrain \textit{all} terms in the low-energy effective Lagrangian, taking into account symmetries and theoretical restrictions \citep{cosmoflow2}. This is particularly important for the collider paradigm, given that several of the tricks used to enhance the oscillatory exchange signal (e.g., reducing the sound-speed \citep{Lee:2016vti,Jazayeri:2022kjy}) amplify the non-oscillatory self-interaction background, motivating a joint treatment.

In this work, we perform such a study, focusing on the EFT of a massive scalar field $\sigma$ coupled to the Goldstone mode $\pi$. Our approach will be twofold: (1) starting from the full cubic Lagrangian, we will compute the combined curvature bispectrum as a function of all relevant quadratic and cubic couplings (including mixing coefficients and sound-speeds); (2) we will place \textit{joint} constraints on these parameters by building a likelihood for the combined model given CMB data from \textit{Planck}. To compute the correlators, we will leverage recent technological developments including the (analytic) inflationary bootstrap \citep{Arkani-Hamed:2018kmz,Baumann:2019oyu,Pajer:2020wnj,Tong:2021wai,Pimentel:2022fsc,Jazayeri:2022kjy,DuasoPueyo:2023kyh,Qin:2022fbv,Qin:2023bjk,Aoki:2023wdc,Xianyu:2023ytd,Qin:2023ejc,Aoki:2024uyi,Liu:2024str,Liu:2024xyi,Jazayeri:2023xcj,Xianyu:2025lbk,Baumann:2026atn,deRham:2025mjh,Pimentel:2025rds,Liu:2024str,Xianyu:2025lbk} and the (numerical) cosmological flow \citep{Mulryne:2009kh,Dias:2016rjq,Pinol:2023oux,Werth:2023pfl,Werth:2024aui,Jazayeri:2023xcj}, with the latter allowing extension to otherwise difficult-to-model regimes (namely, when the mixing between $\pi$ and $\sigma$ becomes large, see also \citep{Pinol:2026xnl,Wang:2026lff,Huenupi:2026aqc}). Typically, application to data is a rate-limiting step: optimal CMB estimators \citep{Komatsu:2003iq} are prohibitively expensive to apply repeatedly (even when combined with extensions to ensure factorizability \citep{Philcox:2025bbo}), and modal estimators \citep{Fergusson:2009nv,Fergusson:2014gea,Sohn:2023fte} can still be rate-limiting when used in sampling pipelines. We overcome this difficulty using the recent binned primordial bispectrum measurements of \citep{Philcox:2025bbo,Philcox:2026njr}, which allow the likelihood to be evaluated in just milliseconds, enabling efficient exploration of the multi-field posterior. 

\vspace{2pt}
The remainder of this work is as follows. In \S\ref{sec: models} we present the multi-field action and the associated inflationary correlation functions, alongside theoretical bounds on the couplings. \S\ref{sec: practical} describes the practical computation of these shapes, as well as the observational dataset and likelihood. In \S\ref{sec: results}, we present the key results: joint constraints on the inflationary coupling amplitudes. We conclude in \S\ref{sec: conclusions} with a summary and a discussion of future work. Throughout, we use the metric signature $(-,+,+,+)$ and compute shape functions in the de Sitter limit (with $n_s=1$). When applying to data, we use the \textit{Planck} baseline cosmology \citep{2020A&A...641A...6P}, whose inflationary sector is specified by $\Delta^2_\zeta(k_{\rm pivot}) \equiv A_s = 2.10\times 10^{-9}$, $k_{\rm pivot}=0.05\;\mathrm{Mpc}^{-1}, n_s=0.966$ and $r=0$.

\section{Inflationary Models \& Shapes}\label{sec: models}
\subsection{The Inflationary Action}
\noindent In the EFT of inflation \citep{Cheung:2007st,Weinberg:2008hq,Senatore:2009gt}, the Lagrangian for the Goldstone mode $\pi$ can be written
\beq
    \mathcal{L}_\pi / a^3 &=& \frac{M_{\rm Pl}^2|\dot H|}{c_\pi^2}\left[ \dot{\pi}^2 - c_{\pi}^2\frac{(\partial_i \pi)^2}{a^2} \right]- \frac{M_{\rm Pl}^2|\dot{H}|}{c_\pi^{2}}(1-c_\pi^2)\left[\frac{(\partial_i \pi)^2}{a^2} \dot{\pi} -\left( 1 + \frac{2}{3} \frac{\tilde{c}_3}{c_\pi^2} \right)\dot{\pi}^3\right]+\cdots,
\eeq
(in the decoupling limit), where $M_{\rm Pl}$ is the Planck mass, $c_\pi\leq 1$ is the sound-speed, $H$ is the Hubble parameter, and $\tilde{c}_3$ is a boost-breaking parameter.\footnote{The $\tilde{c}_3$ term arises from the single-field operator $\mathcal{L}_\pi/a^3 \supset -\tfrac{4}{3}M_3^4\dot{\pi}^3$, with $\tilde{c}_3(c_\pi^{-2}-1)=-2M_3^4c_\pi^2/(M_{\rm Pl}^2|\dot H|)$.} Alongside the quadratic terms, this features two cubic interactions, $(\partial_i\pi)^2\dot\pi$ and $\dot\pi^3$, which encode self-interactions of the inflation. Notably, the amplitude of the first term is fixed by the quadratic theory due to the non-linear realization of boosts \citep{Cheung:2007st}, whilst the second is free (with $\tilde{c}_3=0$ recovering the (almost-)boost-invariant theory). 

In this work, we are interested in the interplay between the Goldstone mode and a massive scalar field, $\sigma$. The latter is defined by the cubic Lagrangian \citep[e.g.,][]{Senatore:2010wk,Pinol:2023oux}
\beq
     \mathcal{L}_\sigma / a^3 &=& \frac{1}{2} \left[ \dot{\sigma}^2 - \frac{(\partial_i \sigma)^2}{a^2} - m_{\sigma}^2\sigma^2 \right] - \mu_{\sigma}\sigma^3,
\eeq
where $\mu_\sigma$ parametrizes leading-order self-interactions. We allow for interactions between $\pi$ and $\sigma$ according to the mixing Lagrangian (as informed by the EFT of inflation):\footnote{These are generated by the coupling of $\sigma$ to the metric perturbations via $\delta g^{00}\sigma$, $(\delta g^{00})^2\sigma$, and $\delta g^{00}\sigma^2$.}
\beq
    \mathcal{L}_{\rm mix} / a^3 &=& 2 \tilde{M}_1^3 \dot{\pi}\sigma-\tilde{M}_1^3\frac{(\partial_i \pi)^2}{a^2} \sigma  + \left( \tilde{M}_1^3 - 4 \tilde{M}_3^3 \right)\dot{\pi}^2 \sigma+2\tilde{M}_2^2\dot{\pi}\sigma^2.
\eeq
Note that the amplitude of the $(\partial_i\pi)^2\sigma$ interaction is fixed by the quadratic mixing term. Finally, we allow for $\sigma$ to have a non-unit sound-speed $c_\sigma\leq 1$ via the operator
\beq
    \mathcal{L}_{c_\sigma}/a^3= \frac{1-c_\sigma^2}{2c_\sigma^2}\left[\dot\sigma^2-2\frac{\partial_i\pi\partial_i\sigma}{a^2}\dot\sigma\right],
\eeq
where the second term arises from non-linearly realized symmetries, and we drop a (small) term in $\dot\sigma\dot \pi^2$ whose amplitude is not fixed by $c_\sigma$.

\begin{figure}
    \centering
    \includegraphics[width=0.9\linewidth]{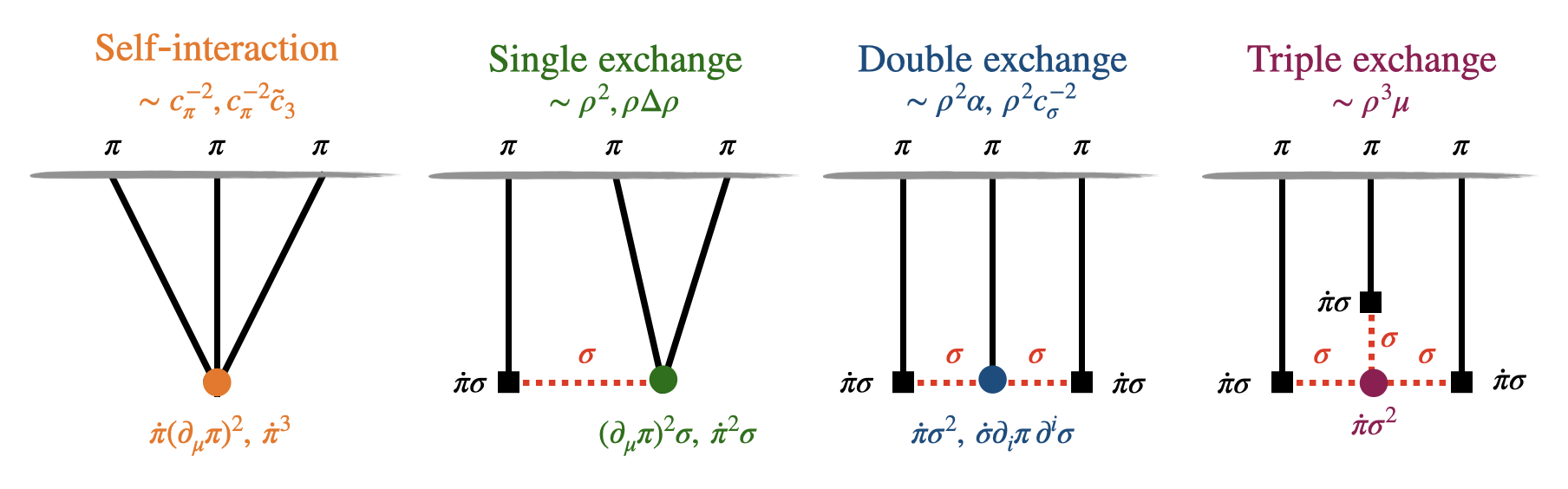}
    \caption{\textbf{Bispectrum Interaction Schematic}: At tree-level, the Lagrangian given in \eqref{eq: lagrangian-cosmoflow} sources four types of interactions: self-interactions of the Goldstone mode, single exchange of a massive scalar, double exchange, and triple exchange. All exchange interactions involve the mixing vertex $\rho\dot\pi\sigma$; for masses $m\gtrsim H$, a detectable signal usually requires $\rho\gtrsim H$, falling in the `strong-mixing' regime (which leads to additional quadratic insertions in the above diagrams). A summary of the physical parameters describing these templates is given in Tab.\,\ref{tab: parameters}.}\label{fig: cartoon}
\end{figure}

The combined Lagrangian, $\mathcal{L} \equiv \mathcal{L}_\pi + \mathcal{L}_\sigma + \mathcal{L}_{\rm mix}$, can be simplified by introducing the canonically normalized fields $\pi_c\equiv f_\pi^2(c_\pi/c_\sigma)^{-3/2}\pi$ and $\sigma_c \equiv c_\sigma^{1/2}\sigma$, for symmetry-breaking scale $f_\pi^4 \equiv 2M_{\rm Pl}^2|\dot{H}|c_\pi$. Additionally redefining $x_i \to c_\sigma x_i$ with $\partial_\mu^2 \equiv a^{-2}c_\sigma^{-2}\partial_i^2-\partial_t^2$, we find
\beq\label{eq: lagrangian-cosmoflow}
    \mathcal{L} / a^3 &=& \frac{1}{2} \left[ \dot{\pi}_c^2 - c_{\rm rel}^2\frac{(\partial_i \pi_c)^2}{a^2} \right]+ \frac{1}{2} \left[ \dot{\sigma}_c^2 - \frac{(\partial_i \sigma_c)^2}{a^2} - m^2 \sigma_c^2 \right] +\rho\dot{\pi}_c \sigma_c\\\nonumber
    &&-\lambda(\partial_\mu\pi_c)^2\dot{\pi}_c-\Delta\lambda\,\dot{\pi}_c^3 -\frac{\kappa}{2}(\partial_\mu\pi_c)^2\sigma_c -\frac{\Delta\kappa}{2}\dot{\pi}_c^2 \sigma_c -\frac{\alpha}{2} \dot{\pi}_c \sigma_c^2 - \mu \sigma_c^3 - \beta\frac{\partial_i\pi_c\partial_i\sigma_c}{a^2}\dot{\sigma}_c
\eeq
\citep[cf.][]{Pinol:2023oux}, defining the coefficients\footnote{In the notation of \citep{Pinol:2023oux,Jazayeri:2023xcj}, $\{\lambda,\Delta\lambda,\kappa,\Delta\kappa\} = \{\lambda_1c_\sigma^2, \lambda_2+\lambda_1c_\sigma^2, \kappa_1c_\sigma^2,\kappa_2+\kappa_1c_\sigma^2\}$, with $\Delta\rho = \rho+\tilde\rho$. We additionally fix a sign error in $\lambda_1$ in \citep{Pinol:2023oux} (noting that $\lambda_1$ and $\lambda_2$ must have opposite signs if $\tilde{c}_3=0$), as well as a sign error and factors of two in the $\kappa_{1,2}$ definitions of \citep{Jazayeri:2023xcj}.}
\beq\label{eq: parameter-mapping}
    c_{\rm rel} &=& \frac{c_\pi}{c_\sigma}, \quad m = c_\sigma m_\sigma, \quad \rho = 2\frac{\tilde{M}_1^3 c_\sigma^{5/2}}{f_\pi^2}c_{\rm rel}^{3/2},\quad \mu = \mu_{\sigma}c_\sigma^{3/2}, \quad \beta = \frac{c_\sigma^{-2}-1}{f_\pi^2}c_{\rm rel}^{3/2}\\\nonumber
    \lambda &=&\frac{1}{2}\frac{(1-c_\pi^2)}{f_\pi^2}c_{\rm rel}^{3/2}, \quad \Delta\lambda = -\frac{1}{3} \frac{\tilde{c}_3}{c_\pi^2}\frac{(1-c_\pi^2)}{f_\pi^2}c_{\rm rel}^{3/2}, \quad \kappa = \frac{c_{\rm rel}^{3/2}}{f_\pi^2}\rho, \quad \Delta\kappa = \frac{c_{\rm rel}^{3/2}}{f_\pi^2}\Delta\rho\\\nonumber
    \Delta\rho &=& 8\frac{\tilde{M}_3^3}{f_\pi^2}c_\sigma^{5/2}c_{\rm rel}^{3/2}, \quad\alpha = -4\frac{\tilde{M}_2^2c_\sigma^2}{f_\pi^2} c_{\rm rel}^{3/2}.
\eeq
The first line of \eqref{eq: lagrangian-cosmoflow} describes the quadratic theory for $\pi_c$ and $\sigma_c$, which depends on three parameters: $c_{\rm rel}\equiv c_\pi/c_\sigma, m, \rho$, where $\rho$ encodes the mixing between sectors (setting $\rho\geq 0$ without loss of generality). As sketched in Fig.\,\ref{fig: cartoon}, the seven terms in the second line generate tree-level bispectra, including self-interactions ($(\partial_\mu\pi_c)^2\dot\pi_c$ and $\dot\pi_c^3$), single exchange ($(\partial_\mu\pi_c)^2\sigma_c$ and $\dot\pi_c^2\sigma_c$), double exchange ($\dot\pi_c\sigma^2$ and $\dot\sigma_c\partial_i\pi_c\partial_i\sigma_c$) and triple exchange ($\sigma^3$) \citep[e.g.,][]{Lee:2016vti,Wang:2022eop,Xianyu:2023ytd,Aoki:2024uyi}. Notably, only four of the seven cubic amplitudes are independent: $\lambda, \kappa,\beta$ are fully determined by $c_\pi,c_\sigma,\rho$ due to the non-linear realization of boosts. In this work, we will parametrize the theory by
\beq
    \{c_\pi, c_\sigma, m, \rho\} + \{\tilde{c}_3, \Delta\rho, \alpha, \mu\},
\eeq
where $\tilde{c}_3$ and $\Delta\rho$ parametrize the boost-breaking interactions $\dot\pi_c^3$ and $\dot\pi_c^2\sigma$.

\subsection{Correlation Functions}
\noindent Given the action, we can the correlation functions of $\pi_c$. Traditionally, this is performed using in-in (Schwinger-Keldysch) methods \citep{Schwinger:1960qe,Keldysh:1964ud,Maldacena:2002vr,Weinberg:2005vy,Chen:2017ryl}; in this work, we will mostly use alternative techniques, as discussed below. The Goldstone mode encodes the curvature fluctuation at the end of inflation via the relation \citep{Maldacena:2002vr,Cheung:2007st}
\beq
    \zeta = -H\pi + H\pi\dot\pi+\frac{1}{2}\dot H\pi^2\cdots
\eeq
Here, only the first term is needed, since the others decay after horizon crossing for sufficiently massive particles. Recalling the definition of $\pi_c$, we can write the curvature power spectrum and bispectrum, $P_\zeta$ and $B_\zeta$, in terms of the correlators of the canonically normalized field:
\beq\label{eq: PzetaBzeta}
    \delD{\vk_1+\vk_2}P_\zeta(k_1) &\equiv& \av{\zeta(\vk_1)\zeta(\vk_2)} = \frac{H^2}{f_\pi^4}c_{\rm rel}^{3}\av{\pi_c(\vk_1)\pi_c(\vk_2)}\\\nonumber
    \delD{\vk_1+\vk_2+\vk_3}B_\zeta(k_1,k_2,k_3) &\equiv& \av{\zeta(\vk_1)\zeta(\vk_2)\zeta(\vk_3)} = -\frac{H^3}{f_\pi^6}c_{\rm rel}^{9/2}\av{\pi_c(\vk_1)\pi_c(\vk_2)\pi_c(\vk_3)}.
\eeq
Motivated by the empirical scale-invariance of the two-point function, we define dimensionless shape functions:\footnote{Many works separate the amplitude and scale information, with $S_\zeta(k_1,k_2,k_3)\to f_{\rm NL}S_\zeta(k_1,k_2,k_3)$, setting $S_\zeta(k,k,k)=1$. We do not perform such a decomposition in this work, since $f_{\rm NL}$ is not well-defined in the strong mixing regime.}
\beq
    \Delta_\zeta^2(k) &=& \frac{k^3}{2\pi^2}P_\zeta(k), \qquad S_\zeta(k_1,k_2,k_3) = \frac{5}{18}(k_1k_2k_3)^2\frac{B_\zeta(k_1,k_2,k_3)}{(2\pi^2\Delta^2_\zeta(k_{\rm pivot}))^2}.
\eeq
In the de Sitter limit (which we shall assume), $\Delta^2_\zeta$ is independent of $k$ and $S_\zeta$ depends only on the ratios $x=k_1/k_3$ and $y=k_2/k_3$. Moreover, in the absence of interactions, $\av{\pi_c\pi_c}\to H^2 / (2c_{\rm rel}^3)$, such that the power spectrum amplitude satisfies $\Delta^2_\zeta = (H/f_\pi)^4/(4\pi^2)\approx 2.1\times 10^{-9}$ \citep{Planck:2018vyg}, giving $f_\pi\approx 59 H$. When interactions are present, 
we use \eqref{eq: PzetaBzeta} to define $f_\pi/H$ given the interaction-dependent two-point function of $\pi_c$; in practice, this leads to an enhancement of $f_\pi$ if $\rho\gtrsim H$.\footnote{Explicitly, we define $f_\pi$ using the best-fit \textit{Planck} amplitude via $4\pi^2\Delta^2_\zeta(f_\pi/H)^4 = \av{\pi_c\pi_c}/\av{\pi_c\pi_c}_{\rm ideal}$.}

\subsection{Phenomenology}\label{subsec: pheno}
\noindent We now discuss the phenomenology of the bispectrum shape function, $S_\zeta$, induced by the effective Lagrangian of \eqref{eq: lagrangian-cosmoflow}. We consider three scenarios depending on the value of $\rho$ relative to the Hubble scale (which will be of relevance to the data analysis in \S\ref{sec: results}): (1) no mixing, $\rho=0$; (2) weak mixing, $\rho\ll H$; (3) strong mixing, $\rho\gtrsim H$. We summarize the various types of interactions and their underlying parameters in Fig.\,\ref{fig: cartoon}\,\&\,Tab.\,\ref{tab: parameters}.

\subsubsection{\texorpdfstring{No Mixing $(\rho=0)$}{No Mixing}}
\noindent For $\rho=0$, $\pi$ and $\sigma$ decouple, thus \eqref{eq: lagrangian-cosmoflow} reduces to the single-field EFT Lagrangian. In this case, non-Gaussianity is generated by the two cubic self-interactions, $(\partial_\mu\pi)^2\dot\pi$ and $\dot\pi^3$, with amplitudes set by $\lambda$ and $\Delta\lambda$. Explicitly, the tree-level bispectrum is given by
\beq\label{eq: no-mixing}
    S_\zeta^{\rm no}(x,y;c_\pi,\tilde{c}_3) &=& \lambda(c_\pi,c_\sigma)S_\zeta^{(\lambda)}(x,y;c_{\rm rel})+\Delta\lambda(c_\pi,c_\sigma,\tilde{c}_3) S_\zeta^{(\Delta\lambda)}(x,y;c_{\rm rel}),
\eeq
where $S_\zeta^{(\theta)}$ indicates the shape evaluated at unit $\theta$ setting all other cubic couplings to zero. Note that the templates depend non-trivially on parameters entering the quadratic Lagrangian. Using the in-in formalism, the underlying templates can be
computed explicitly \citep[e.g.,][]{Maldacena:2002vr,Chen:2006nt,Senatore:2009gt}, yielding
\beq\label{eq: no-mixing-templates}
    S^{(\lambda)}_\zeta(x,y;c_{\rm rel}) &=& \frac{5f_\pi^2}{36}c_{\rm rel}^{-7/2}\frac{24K_3^6-8K_2^2K_3^3K_1-8K_2^4K_1^2+22K_3^3K_1^3-6K_2^2K_1^4+2K_1^6}{K_3^9K_1^3}\\\nonumber
    S^{(\Delta\lambda)}_\zeta(x,y;c_{\rm rel}) &=& \frac{10f_\pi^2}{3}c_{\rm rel}^{-3/2}\frac{1}{K_1^3K_3^3}
\eeq
defining $K_1=k_1+k_2+k_3, K_2=(k_1k_2+k_2k_3+k_3k_1)^{1/2}, K_3=(k_1k_2k_3)^{1/3}$ with $x = k_1/k_3, y=k_2/k_3$ as before. As expected, the dependence on $c_\sigma$ cancels in the overall shape such that $S_\zeta^{\rm no}$ depends only on $c_\pi$ and $\tilde{c}_3$ (cf.\,Tab.\,\ref{tab: parameters}). 
Inserting the definitions of $\lambda$ and $\Delta\lambda$ \eqref{eq: parameter-mapping}, we find that the bispectra sourced by the $(\partial_\mu\pi)^2\dot\pi$ and $\dot\pi^3$ interactions scale as $c_\pi^{-2}$ and $\tilde{c}_3(c_\pi^{-2}-1)$ in the small sound-speed limit; these combinations can be directly constrained from observational data \citep[cf.][]{Creminelli:2005hu,Senatore:2009gt}.

\begin{table}[]
    \centering
    \begin{tabular}{c|l|lll}
    \textbf{Parameter} & \textbf{Description} & No? & Weak? & Strong?\\\hline
     $c_\pi$ & Goldstone sound-speed & \checkmark & \checkmark & \checkmark\\
     $c_\sigma$ & Scalar sound-speed & -- & \checkmark & \checkmark\\ 
     $m$ & Scalar mass & -- & \checkmark & \checkmark\\ 
     $\rho$ & Quadratic mixing & -- & \checkmark & \checkmark\\ \hline
     $\tilde{c}_3$ & $\dot\pi^3$ self-interaction  & \checkmark & \checkmark & \checkmark\\
     $\Delta\rho$ & $\dot\pi^2\sigma$ single exchange  & -- & \checkmark & \checkmark\\
     $\alpha$ & $\dot\pi\sigma^2$ double exchange  & -- & -- & \checkmark\\
     $\mu$ & $\sigma^3$ triple exchange  & -- & -- & \checkmark\\
    \end{tabular}
    \caption{\textbf{Inflationary Lagrangian Parameters}: List of underlying parameters describing the propagation and interactions of the inflationary Goldstone mode $\pi$ and the massive scalar $\sigma$. For each parameter, we list which part of the bispectrum it controls, and indicate whether it enters the unmixed, weakly-mixed, and strong-mixed scenarios. These enter the two-field Lagrangian \eqref{eq: lagrangian-cosmoflow} via \eqref{eq: parameter-mapping}, and are subject to various perturbativity and unitarity constraints (\S\ref{subsec: bounds}).}
    \label{tab: parameters}
\end{table}

\subsubsection{\texorpdfstring{Weak Mixing $(\rho\ll H)$}{Weak Mixing}}
\noindent Next, we allow for a small mixing of the Goldstone mode with the massive scalar, $\sigma$. Since the double- and triple exchange diagrams shown in Fig.\,\ref{fig: cartoon} require additional $\dot\pi\sigma$ mixings (each of which is suppressed by $\rho/H$), we focus on the single exchange diagrams arising from $(\partial_\mu\pi)^2\sigma$ and $\dot\pi^2\sigma$, whose amplitudes scale as $\rho\kappa\propto \rho^2$ and $\rho\Delta\kappa\propto \rho\Delta\rho$ respectively (following \citep[e.g.,][]{Lee:2016vti,Pimentel:2022fsc,Sohn:2024xzd,cosmoflow2}). Including the self-interaction terms from \eqref{eq: no-mixing}, we obtain the bispectrum shape
\beq\label{eq: weak-exchange}
    S_\zeta^{\rm weak}(x,y;c_\pi,c_\sigma,m,\rho,\tilde{c}_3,\Delta\rho) &=& \lambda(c_\pi,c_\sigma)S_\zeta^{(\lambda)}(x,y;c_{\rm rel})+\Delta\lambda(c_\pi,c_\sigma,\tilde{c}_3) S_\zeta^{(\Delta\lambda)}(x,y;c_{\rm rel})\\\nonumber
    &&\,+\,\rho\,\kappa(\rho,c_\pi,c_\sigma)S^{(\rho\kappa)}_\zeta(x,y;c_{\rm rel},m)+\rho\,\Delta\kappa(\rho,\Delta\rho,c_\pi,c_\sigma)S^{(\rho\Delta\kappa)}_\zeta(x,y;c_{\rm rel},m).
\eeq
As for the $\rho=0$ case, the combined bispectrum factorizes into a product of amplitudes, e.g., $\rho\kappa$, depending on quadratic and cubic exchange parameters, and templates, e.g., $S_\zeta^{(\rho\kappa)}$, depending only on $m$ and $c_{\rm rel}\equiv c_\pi/c_\sigma$.\footnote{We define $S^{(\rho\kappa)}_\zeta \equiv \left.\partial S_\zeta/\partial (\rho\kappa)\right|_{\rho=\lambda=\Delta\lambda=\Delta\kappa=0}$, working in the $\rho\to0$ limit.}

In contrast to the self-interaction templates, $S_\zeta^{(\lambda,\,\Delta\lambda)}$, the single exchange templates, $S_\zeta^{(\rho\kappa,\,\rho\Delta\kappa)}$, do not have a simple analytic form. As discussed in \citep{Lee:2016vti}, they can be represented as the sum of two components: (1) an analytic piece, which dominates in the equilateral limit ($x\approx y\approx 1$) and scales as $\mu(m)^{-2}$ (for $\mu(m) = \sqrt{m^2/H^2-9/4}$); (2) a non-analytic piece, with the asymptotic form
\beq\label{eq: weak-squeezed}
    S_\zeta(x\ll y) \sim x^{1/2+i\mu(m)} + \text{c.c.};
\eeq
sourcing oscillations in $\log x$ with frequency $\mu(m)$ if $m\geq 3H/2$ \citep{Chen:2009zp,Noumi:2012vr,Arkani-Hamed:2015bza,Lee:2016vti}. In practice, the latter signatures are suppressed by $e^{-\pi\mu(m)}$, making them difficult to detect if $m/H$ is large. For $c_\pi\ll 1$, the $(\partial_\mu\pi)^2\sigma$ terms are enhanced by a factor $c_{\pi}^{-2}$, and the oscillations instead scale as $e^{-\pi\mu(m)/2}$; we caution that this boost in signal is accompanied by a large self-interaction background. 

\subsubsection{\texorpdfstring{Strong Mixing $(\rho\gtrsim H)$}{Strong Mixing}}\label{subsec: strong-theory}
\noindent Finally, we consider the full tree-level bispectrum, incorporating all terms in the Lagrangian of \eqref{eq: lagrangian-cosmoflow} and allowing for arbitrary values of the mixing coefficient $\rho$.\footnote{In this work, we use the term `strong mixing' to refer to arbitrary $\rho$, not just the asymptotic regime discussed in \citep{Pinol:2023oux,Jazayeri:2023xcj} (which requires $\rho/H\gtrsim c_s^{-1}$ and $c_sm^2/\rho \lesssim H$).} The combined bispectrum shape is the sum of two self-interactions, two single exchanges, two double exchanges and one triple exchange (see Fig.\,\ref{fig: cartoon}), each of which can be written in terms of a microphysical amplitude and a macrophysical template:
\beq\label{eq: strong-exchange}
    S_\zeta^{\rm strong}(x,y;c_\pi,c_\sigma,m,\rho,\tilde{c}_3,\tilde\rho, \alpha, \mu) &=& \lambda(c_\pi,c_\sigma)S_\zeta^{(\lambda)}(x,y;\Theta_2)+\Delta\lambda(c_\pi,c_\sigma,\tilde{c}_3)S_\zeta^{(\Delta\lambda)}(x,y;\Theta_2)\\\nonumber
    &&\,+\,  \kappa(\rho,c_\pi,c_\sigma)S_\zeta^{(\kappa)}(x,y;\Theta_2)+\Delta\kappa(\tilde{\rho},c_\pi,c_\sigma)S_\zeta^{(\Delta\kappa)}(x,y;\Theta_2)\\\nonumber
    &&\,+\,\alpha\,S_\zeta^{(\alpha)}(x,y;\Theta_2)+\beta(c_\pi,c_\sigma)S_\zeta^{(\beta)}(x,y;\Theta_2)\\\nonumber
    &&\,+\,\mu\,S_\zeta^{(\mu)}(x,y;\Theta_2),
\eeq
where $\Theta_2 \equiv\{c_{\rm rel}, m, \rho\}$. This has a number of differences from \eqref{eq: weak-exchange}: (1) all templates depend explicitly on $\rho$ and $m$ since the mixing alters the propagation of the Goldstone mode;\footnote{Diagramatically, this corresponds to additional $\pi$-$\sigma$ mixings in Fig.\,\ref{fig: cartoon}.} (2) the single exchange diagrams scale as $\rho\kappa_i$ only for $\rho\lesssim H$ (and similarly for double- and triple exchange diagrams); (3) the self-interaction signatures are modified by the exchange; (4) the power spectrum, $\av{\zeta^2}$, gains explicit dependence on $\Theta_2$, which impacts the conversion from $\pi_c$ to $\zeta$ (cf.\,\ref{eq: PzetaBzeta}). 

As discussed in \citep{Pinol:2023oux,Jazayeri:2023xcj,cosmoflow2,Pinol:2026xnl,Wang:2026lff,Huenupi:2026aqc}, the phenomenology of $S_\zeta^{\rm strong}$ is determined by a complex interplay of $\rho$, $m$, and $c_{\rm rel}$, with the exchange and self-interaction contributions inexorably intertwined for $\rho\gtrsim H$. Though the full shape must be computed numerically, the squeezed limit takes a similar form to \eqref{eq: weak-squeezed}:
\beq\label{eq: strong-squeezed}
    S_\zeta(x\ll y) \sim x^{1/2+i\mu_{\rm eff}} + \text{c.c.} \quad \text{for} \quad \mu_{\rm eff} = \sqrt{(m^2+\rho^2)/H^2-9/4};
\eeq
with oscillations if the effective mass, $\sqrt{m^2+\rho^2}$, is larger than $3H/2$. Analogously to the weakly-mixed case, the amplitude scales as $e^{-\pi\mu_{\rm eff}}$, implying that regimes with $\rho\gtrsim H$ will be difficult to detect in practice. Finally, we note that the massive field $\sigma$ can be integrated out in the limit of $\rho\gg \mathrm{max}(m, H)$, transforming the two-field system into a non-local one-field model \citep{Jazayeri:2023xcj}. This has the effective Lagrangian
\beq\label{eq: lagrangian-cosmoflow-nonlocal}
    \mathcal{L}_{\rm non\text{-}local} / a^3 &=& \frac{1}{2} \left[ \dot{\pi}_c\left(1+\rho^2\mathcal{D}^{-1}\right)\dot{\pi}_c - c_{\rm rel}^2\frac{(\partial_\mu \pi_c)^2}{a^2} \right]-\lambda \frac{(\partial_\mu \pi_c)^2}{a^2} \dot{\pi}_c -\Delta\lambda  \dot{\pi}_c^3\\\nonumber
    && -\frac{\kappa\rho}{2} \frac{(\partial_\mu \pi_c)^2}{a^2}\mathcal{D}^{-1}\dot{\pi}_c -\frac{\kappa_2\rho}{2}\dot{\pi}_c^2 \mathcal{D}^{-1}\dot{\pi}_c -\frac{\alpha\rho^2}{2} \dot{\pi}_c \left(\mathcal{D}^{-1}\dot{\pi}_c\right)^2 - \mu\rho^3\left(\mathcal{D}^{-1}\dot{\pi}_c\right)^3,
\eeq
where $\mathcal{D}^{-1} = (-\partial_i^2/a^2+m^2)^{-1}$ and we have dropped the $\dot\sigma\partial_i\pi\partial_i\sigma$ term, which becomes higher-order in derivatives.\footnote{This is obtained by fixing $\sigma_c$ to the saddle-point solution $\sigma_c = \rho\mathcal{D}^{-1}\dot{\pi}_c$.} Using \eqref{eq: lagrangian-cosmoflow-nonlocal}, we can compute the bispectrum shape analogously to before: this provides an excellent approximation for $\rho\gg \mathrm{max}(m,H)$, though does not capture the (exponentially suppressed) oscillations in the squeezed limit. 

\subsection{Theoretical Amplitude Bounds}\label{subsec: bounds}
\noindent The parameters controlling the tree-level inflationary bispectrum (see Tab.\,\ref{tab: parameters}) are subject to a number of theoretical constraints, which ensure that the theory is perturbative, unitary, and self-consistent. In the unmixed theory, we require $c_\pi\geq (2\pi\Delta^2_{\zeta})^{1/4}$ to avoid strong-coupling in the inflaton sector (which would indicate a breakdown of the EFT expansion); this practically restricts us to $c_\pi\gtrsim 0.01$ \citep{Cheung:2007st,Baumann:2011su}. A similar argument leads to the constraint $\tilde{c}_3(c_\pi^{-2}-1) \lesssim (2\pi\Delta_\zeta^2)^{-1/2}\approx 10^4$ \citep{Senatore:2009gt}.

Weak mixing requires that the quadratic interaction, $\rho\dot\pi\sigma$, constitutes only a small correction to the free Lagrangian of the Goldstone mode and massive field (cf.\,\ref{eq: lagrangian-cosmoflow}). Additionally requiring that the cubic interactions are subdominant leads to the approximate parameter bounds (including also the double- and triple exchange diagrams)
\beq\label{eq: bounds-weak}
    \frac{\rho}{H}\lesssim \mathrm{min}\left(c_{\rm rel}^{-1/2}g(c_{\rm rel}), \frac{m}{H}\right), \quad \left|\frac{\Delta\rho}{H}\right|\lesssim \frac{c_{\rm rel}^{-1/2}g(c_{\rm rel})}{2\pi\Delta_\zeta}, \quad |\alpha| \lesssim c_{\rm rel}^{1/2}g(c_{\rm rel}), \quad \left|\frac{\mu}{H}\right|\lesssim g(c_{\rm rel})
\eeq
\citep[cf.][]{Pinol:2023oux}, where $g(c_{\rm rel}) \equiv \mathrm{min}(1,c_{\rm rel}^{-1})$ accounts for the fact that the kinetic term of $\sigma$ is subdominant to that of $\pi$ if $c_\pi>c_\sigma$. Note that the bound on $\Delta\rho$ is around $3500$ times weaker than that on $\rho$. As shown in \citep{cosmoflow2}, these place severe restrictions on the weak mixing parameter space.


In the strong mixing regime, the constraints on all parameters weaken considerably. From \citep{Pinol:2023oux}, we have the approximate bounds\footnote{For $\rho/H$, we use the limit computed in \citep{Jazayeri:2022kjy}, which accounts for factors of $c_{\rm rel}$ appearing in the $f_\pi/H$ definition. In contrast to previous works, we also allow for non-unit $c_\sigma$.}
\beq\label{eq: bounds-strong}
    \frac{\rho}{H} \lesssim \frac{c_\sigma^2c_{\rm rel}\kappa^{1/2}}{\Delta_\zeta}, \quad \left|\frac{\Delta\rho}{H}\right|\lesssim \frac{c_{\rm rel}^{-1/4}}{2\pi \Delta_\zeta}\left(\frac{\rho}{H}\right)^{3/4},
    \quad |\alpha|\lesssim c_{\rm rel}^{1/4}\left(\frac{\rho}{H}\right)^{3/4}, \quad \left|\frac{\mu}{H}\right|\lesssim c_{\mathrm{rel}}^{-3/4}\left(\frac{\rho}{H}\right)^{3/4} 
\eeq
where $\kappa \equiv 2\Gamma(5/4)^2/\pi^3\approx 0.053$. Numerically, this corresponds to $\rho/H \lesssim 5000,\,\left|\Delta\rho/H\right|\lesssim 2\times 10^6,\,  |\alpha|\lesssim 600,\,\left|\mu/H\right|\lesssim 1400$, setting $c_\sigma=c_\pi=1$ with $\rho=\rho_{\rm max}$. Note that \citep{Kumar:2026dih} imposed the tighter limit $|\alpha| \lesssim 10^{-4}$, assuming that both $\dot\pi^2\sigma$ and $\dot\pi\sigma^2$ were generated from the same operator. 
Though the space of allowed couplings is much larger than for weak mixing, we caution that the desired `cosmological collider' oscillations are highly suppressed for large $\rho$, and the overall amplitudes are suppressed at large $\rho$ due to the modified power spectra.

The above bounds do not guarantee that the theory is technically natural (\textit{i.e.}\ stable under radiative corrections). Imposing that the scalar mass, $m$, is not significantly altered by loop effects gives the rough bounds for $c_\sigma=1$:
\beq\label{eq: naturalness}
    \frac{\rho}{H} \lesssim 1, \qquad \left|\frac{\Delta\rho}{H}\right|\lesssim c_{\pi}^{-2}, \qquad |\alpha|\lesssim c_{\pi}^{1/2}(2\pi\Delta_\zeta)^{1/2}, \qquad \left|\frac{\mu}{H}\right|\lesssim 1,
\eeq
where the first restricts to the weak mixing regime. As we see in \S\ref{sec: results}, these constraints are much harder to satisfy in practice (particularly for $\alpha$), though this is not necessarily a limitation of our approach.

\section{Practical Computation}\label{sec: practical}
\subsection{Inflationary Templates}\label{subsec: templates}
\noindent The inflationary shape functions given in \eqref{eq: weak-exchange}\,\&\,\eqref{eq: strong-exchange} depend on a set of scaling amplitudes (set by quadratic and cubic couplings), and dimensionless templates (set by quadratic parameters). These templates can be computed using a variety of techniques, including direct `in-in' integration \citep{Maldacena:2002vr,Weinberg:2005vy,Chen:2009zp,Chen:2017ryl} (including extension to coupled mode-functions \citep{An:2017hlx,Kumar:2026dih,Kumar:2026ogn,Huenupi:2026abj,Huenupi:2026aqc,Pinol:2026xnl}), symmetry-based bootstrap techniques \citep{Arkani-Hamed:2018kmz,Baumann:2019oyu,Pajer:2020wnj,Baumann:2022jpr,Pimentel:2022fsc,DuasoPueyo:2023kyh,Qin:2023ejc,Xianyu:2023ytd,deRham:2025mjh}, spectral decompositions \citep{Sleight:2019hfp,Qin:2022fbv,Xianyu:2022jwk,Liu:2024xyi,Werth:2024mjg}, numerical flow-based techniques \citep{Mulryne:2009kh,Dias:2016rjq,Pinol:2023oux,Werth:2023pfl}, and beyond. Notably, many methods are restricted to weak mixing to avoid having to resum an infinite set of quadratic insertions; this represents an important theoretical limitation. Here, we will adopt two modeling approaches, which are appropriate for the weak and strong mixing limits. Our goal is to predict each template $S_\zeta^{(X)}(x,y;\Theta_2)$ for a set of $x,y$ points on a grid of quadratic parameters $\Theta_2$, which will allow the combined shapes to be compared to observational data.

\begin{figure}
    \centering
    \includegraphics[width=\linewidth]{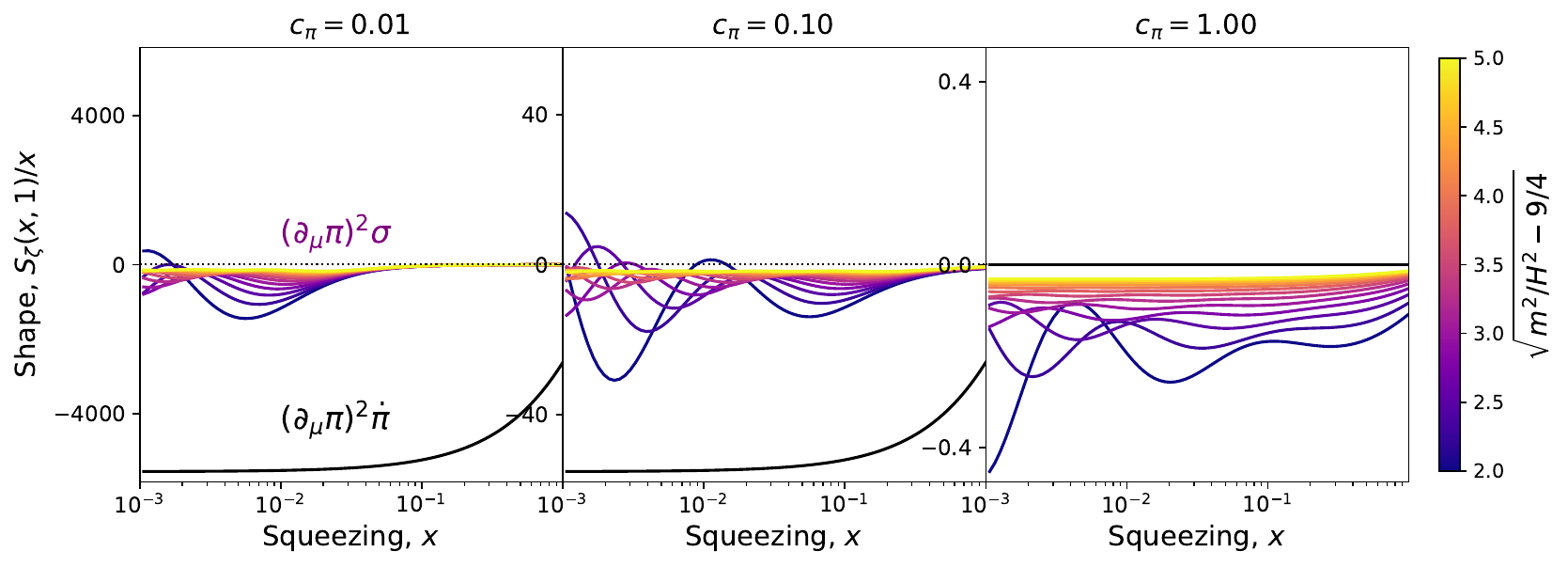}
    \includegraphics[width=\linewidth]{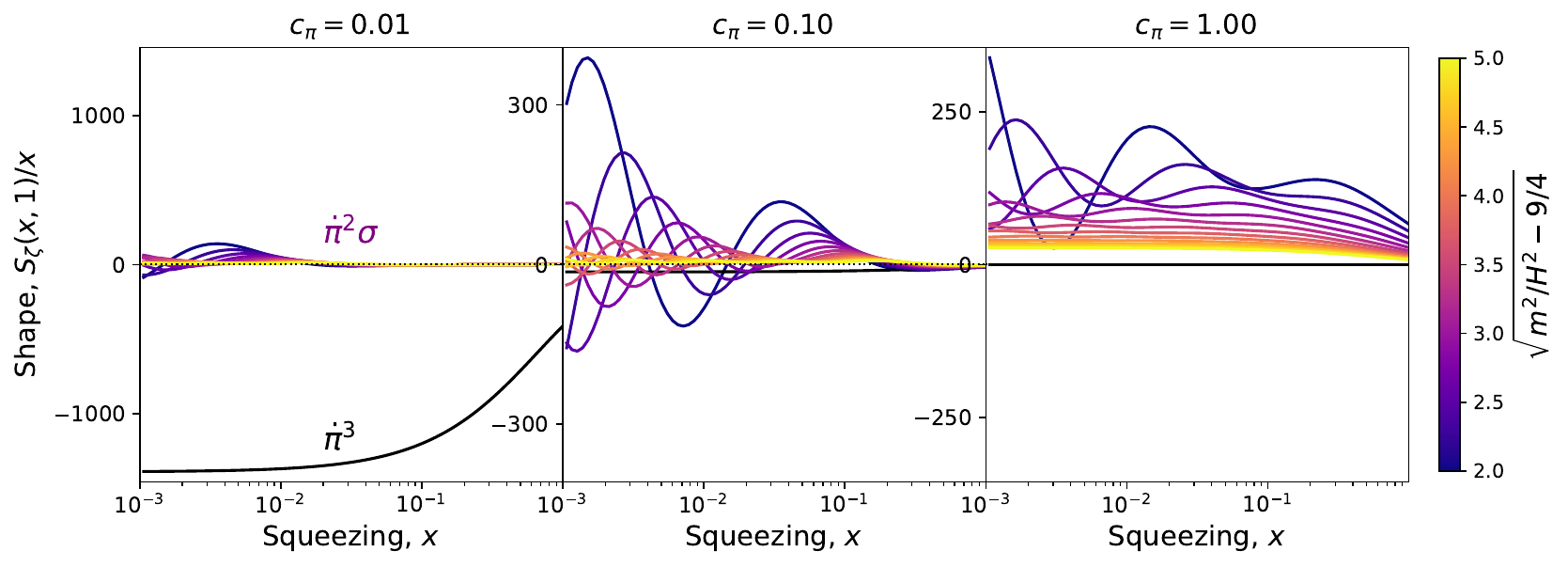}
    \caption{\textbf{Weak Mixing Bispectra}: Isosceles slice of the bispectrum shape function, $S_\zeta(x, 1)$, for various choices of exchange mass $m$ and sound-speed $c_\pi$ (fixing $c_\sigma=1$). We include contributions from both single-exchange (colors) and self-interactions (black). The top panels show contributions generated non-linearly from the quadratic theory (set by $\rho^2$ and $c_\pi^{-2}$), while the bottom give those from boost-breaking interactions (set by $\rho\Delta\rho$ and $(1-c_\pi^{-2})\tilde{c}_3$). Note that we restrict to $m\geq 2.5H$, where the collider oscillations are most apparent. To maximize the signals, we set $\rho=H$, $\Delta\rho=10^3H$ and $\tilde{c}_3=1$.
    Notably, the $\rho^2$ contributions are boosted by $c_\pi\to 0$; this comes at the expense of a large self-interaction background. Furthermore, the exchange signals, including the characteristic cosmological collider oscillations, are highly suppressed for $m\gg H$. An interactive version of this plot can be found \href{https://oliverphilcox.github.io/cosmological-collider-bispectra/}{online}.}
    \label{fig: weak-shapes}
\end{figure}

\subsubsection{\texorpdfstring{$\rho\ll H$: Cosmological Bootstrap}{Cosmological Bootstrap}}
\noindent Assuming weak mixing, the bispectrum is the sum of two self-interactions (which have simple analytic forms, cf.\,\ref{eq: no-mixing-templates}) and two single exchange diagrams. To compute the latter, we utilize the cosmological bootstrap following \citep{Pimentel:2022fsc} (supplemented with the modifications of \citep{Cabass:2024wob}, which allow for $c_{\rm rel}>1$, see also \citep{Jazayeri:2022kjy,Wang:2022eop}). Starting from a set of symmetry assumptions, this computes the three-point function of a conformally-coupled scalar field by solving boundary differential equations \citep{Arkani-Hamed:2018kmz} (in contrast to the traditional integration-over-the-bulk approach). The three-point function of the Goldstone mode can then be computed using `weight-shifting' derivative operators \citep{Arkani-Hamed:2018kmz,Baumann:2019oyu}, whose structure is set by the vertex in question ($(\partial_\mu\pi)^2\sigma$ or $\dot\pi^3\sigma$). The net result is an analytic form for $S_\zeta^{(\rho\kappa_i)}$, which can be represented using series expansions and hypergeometric functions.

Following the pipeline developed for \citep{Philcox:2026njr}, we implement the bootstrap computation in \textsc{mathematica}, computing the two exchange shapes for 171 $(x,y)$ pairs (see \S\ref{subsec: data}), across $81$ logarithmically-spaced points in $c_{\rm rel}\in[10^{-2},10^2]$. Furthermore, we evaluate the model for $66$ choices of exchange mass, $m$, across both the complementary series ($0<m\leq 3H/2$, with power-law behavior as $x\to 0$) and the principal series ($m>3H/2$, with oscillations as $x\to 0$).\footnote{Explicitly, we use $15$ linearly-spaced points with $\sqrt{9/4-m^2/H^2}\in[-1.49,-0.09]$ and $51$ with $\sqrt{m^2/H^2-9/4}\in[0,5]$.} The full computation required approximately 2000 CPU-hours. 

In Fig.\,\ref{fig: weak-shapes} we show an isosceles slice of the shape functions for each of the four interactions discussed above. A number of features are of relevance:
\begin{itemize}
    \item The exchange signals contain oscillations in the squeezed limit, as in \eqref{eq: weak-squeezed}. For large $m$, these are restricted to highly squeezed configurations (which give minimal contributions to the signal-to-noise in the full $(x,y)$-plane), and the shape function approaches a self-interaction-like form. 
    \item To obtain a large signal in the $(\partial_\mu\pi)^2\sigma$ channel, we require either large $\rho/H$ or small $c_\pi$. Assuming weak mixing, $\rho/H$ must be less than order unity, thus a detectable signal requires $c_\pi\ll 1$. Such signals are necessarily accompanied by large self-interaction backgrounds (from $(\partial_\mu\pi)^2\dot\pi$), which limit detectability.
    \item The boost-breaking exchange signal ($\dot\pi^2\sigma$) exhibits only mild sensitivity to $c_\pi$, and, for sufficiently large $\rho\Delta\rho$, can be large. However, both $\Delta\rho$ and $\rho$ are restricted by perturbativity considerations (\S\ref{subsec: bounds}); moreover, at large $m/H$, the exchange signals are degenerate with the self-interactions from $\dot\pi^3$, whose amplitude is unconstrained by the quadratic theory.
\end{itemize}
The full bispectrum combines all four contributions, resulting in a complex non-linear system; we will explore the joint space of quadratic and cubic couplings in \S\ref{sec: results} using Markov Chain analyses. To aid interpretability, we provide an interactive visualization of the weak mixing bispectrum, including all self-interaction and single exchange contributions online.\footnote{Available at \href{https://oliverphilcox.github.io/cosmological-collider-bispectra/}{https://oliverphilcox.github.io/cosmological-collider-bispectra/}.}

\subsubsection{\texorpdfstring{$\rho\gtrsim H$: Cosmological Flow}{Cosmological Flow}}
\noindent Though the cosmological bootstrap provides an efficient way to obtain analytic predictions for inflationary bispectra, it has certain limitations: the approach assumes weak mixing (such that the dependence on $\rho$ factorizes), and practical predictions are not yet available for all diagrams (though see \citep{Aoki:2024uyi,Liu:2024str,Xianyu:2025lbk} for progress in this direction). To analyze the more general scenario with (potentially) large mixing, we use the Cosmological Flow approach described in \citep{Pinol:2023oux,Werth:2023pfl} (see also \citep{Mulryne:2009kh,Dias:2016rjq,Huenupi:2026aqc,Pinol:2026xnl,Wang:2026lff}), which obtains the correlation functions of $\pi_c$ by numerically solving differential equations for their time evolution from Bunch-Davies initialization \citep{Bunch:1978yq} to horizon crossing. This method treats the quadratic Lagrangian non-perturbatively (though remains perturbative in cubic interactions), and can thus be robustly applied to both weak and strong mixing regimes.\footnote{See \citep{Kumar:2026dih,Kumar:2026ogn} for an alternative approach using the coupled mode function method \citep{An:2017hlx}.}

\begin{figure}
    \centering
    \includegraphics[width=0.9\linewidth]{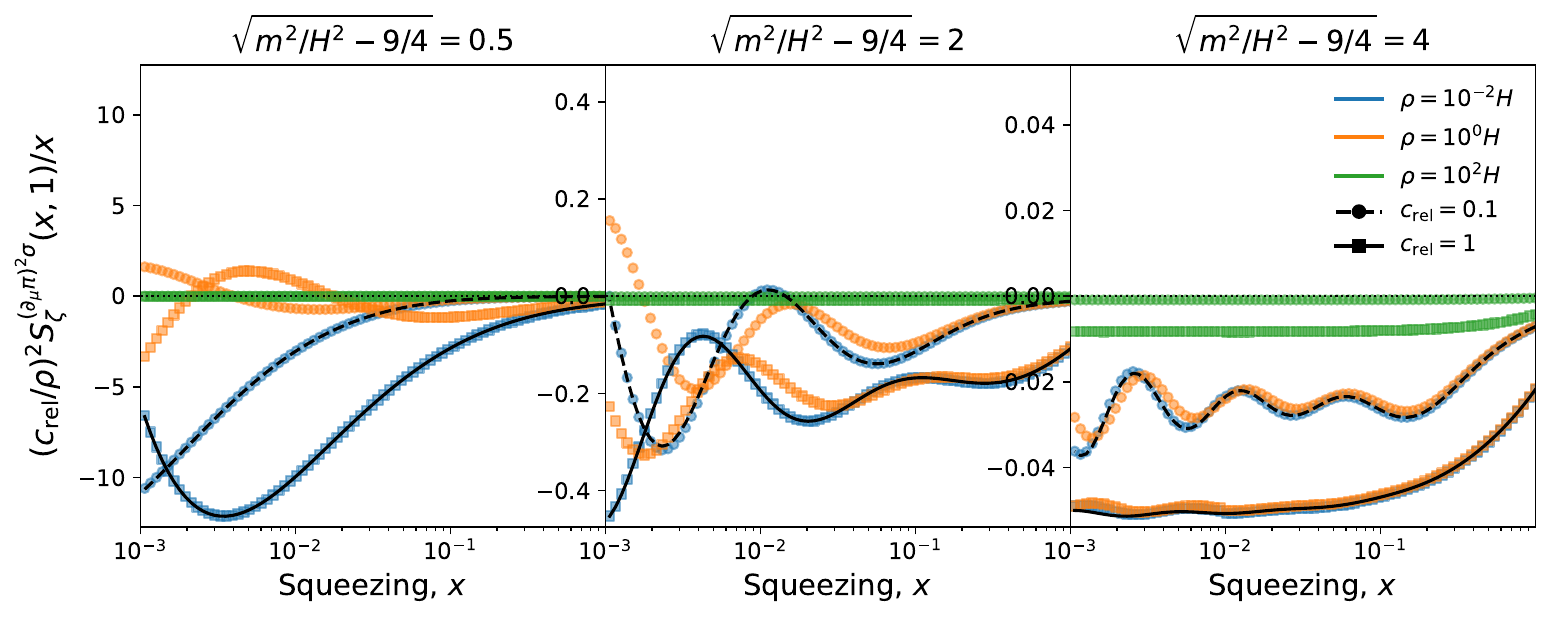}
    \caption{\textbf{Comparison of Bootstrap and \textsc{CosmoFlow} Bispectra}: Shape functions for the $(\partial_\mu\pi)^2\sigma$ interaction computed using bootstrap (black) and \textsc{CosmoFlow} (colors) methods for various masses and sound-speeds. The normalization is chosen such that the bootstrap predictions (which assume weak mixing) are independent of $\rho$. For $\rho = 0.01H$, we find excellent agreement between the two methods. As the mixing grows, the amplitude of the \textsc{CosmoFlow} shape decreases relative to the $\rho^2$ prediction and both the oscillation frequency and damping increases. This suggests that it will be difficult to distinguish particle exchange and self-interactions at large $\rho$.}
    \label{fig: strong-shape-comparison}
\end{figure}

Using the \textsc{CosmoFlow} code \citep{Werth:2024aui}, we compute the seven shape templates entering \eqref{eq: strong-exchange} for $171$ $(x,y)$-points across a broad grid of $\Theta_2 = \{c_{\rm rel}, m, \rho\}$ parameters (in addition to the two-point function, which is needed to calibrate $f_\pi$ via \eqref{eq: PzetaBzeta}). We make a number of modifications to the code: (1) we add the $\beta\dot\sigma\partial_i\sigma\partial_i\pi$ coupling; (2) we \textsc{cythonize} the rate-limiting steps of the algorithm to expedite performance; (3) we replace the hyperbolic tangent adiabatic turn-on function with an error function to reduce transient oscillations.\footnote{We also fix a small error in the flow equation definitions used in \citep{Pinol:2023oux,Werth:2024aui}, which leads to a sign error in the self-exchange shapes. In particular, we replace $3D^{\bar\alpha}_{\beta\gamma}$ with $-3D^{\bar\alpha}_{\beta\gamma}$ in the right-hand side of (3.12) of \citep{Pinol:2023oux} (and (2.9) of \citep{Werth:2024aui}).} Being a numerical solver, the method comes with a number of precision parameters: following extensive testing, we fix to an absolute integration tolerance of $10^{-100}$,\footnote{Close to the folded regime ($x+y\approx 1$), this value is not attainable; in this case, we gradually reduce the tolerance until the numerical solver completes successfully.} a relative tolerance of $10^{-8}$, and start the integration $\Delta N$ $e$-folds before horizon crossing, with $\Delta N\in [6,10]$ (set adaptively from $c_{\rm rel}$ and $\rho$, and iterated until convergence is achieved).

\begin{figure}
    \centering
    \includegraphics[width=0.9\linewidth]{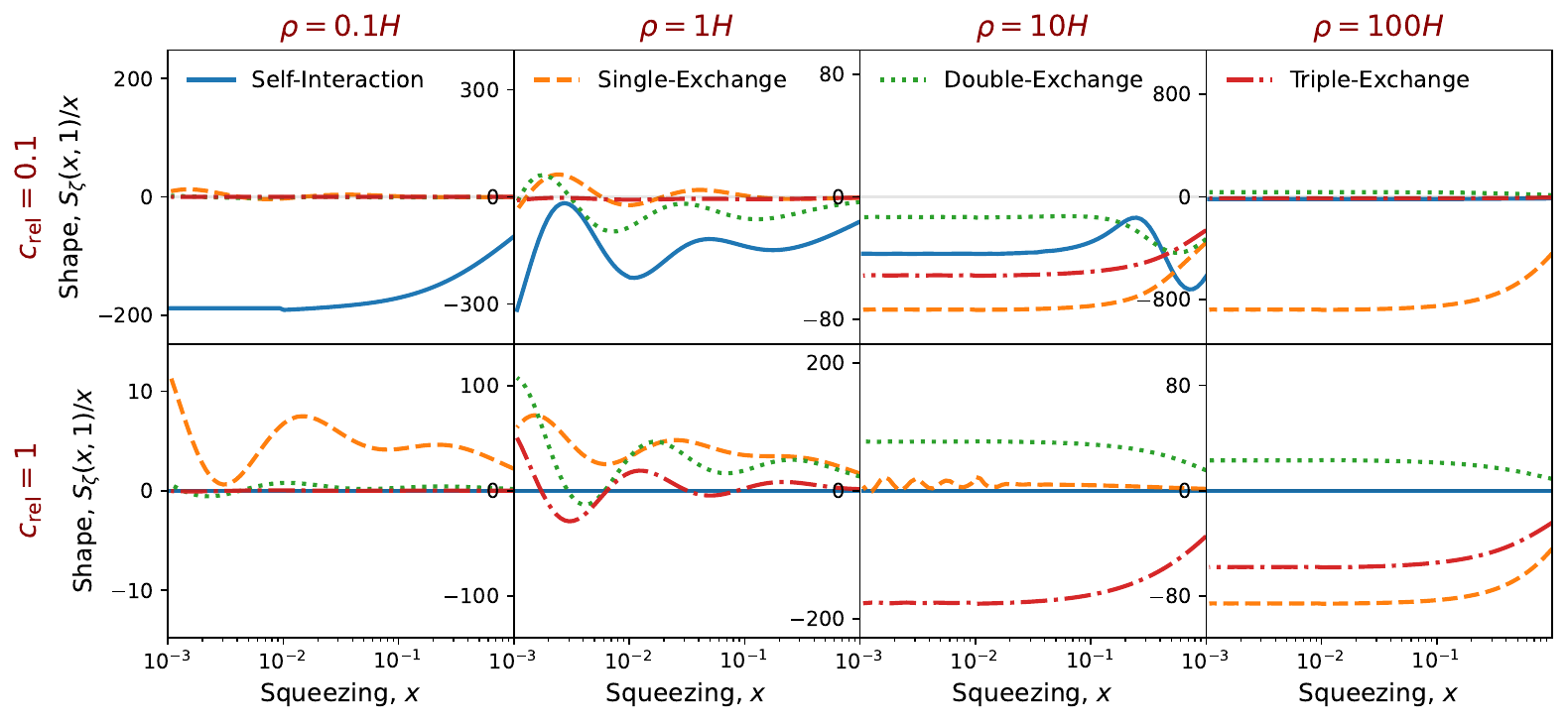}
    \caption{\textbf{Strong Mixing Shapes}: Bispectrum shape functions for the four types of interaction sketched in Fig.~\ref{fig: cartoon}, setting $m=2.5H$. All shapes are computed numerically using \textsc{CosmoFlow} with $c_\sigma=1$, and we restrict to isosceles configurations. Motivated by the theory bounds in \S\ref{subsec: bounds}, we fix $\tilde{c}_3 = 10$, $\Delta\rho=10^3H$, $\alpha=1$ and $\mu = H$. The various shapes depend non-trivially on both the mixing amplitude and sound-speed, with the terms generated by $(\partial_\mu\pi)^2\dot\pi$ and $\dot\pi^3$ (labeled self-interactions) developing oscillations at $\rho/H\sim 1$. At large $\rho/H$, the `cosmological collider' features are lost, and all diagrams asymptote to self-interaction-like shapes. We provide an interactive version of this plot \href{https://oliverphilcox.github.io/cosmological-collider-bispectra/}{online}.}
    \label{fig: strong-all-shapes}
\end{figure}

We evaluate the theoretical models across logarithmically-spaced grids of $c_{\rm rel}\in[10^{-2},10]$ (with spacing $0.5\,\mathrm{dex}$) and $\rho\in[10^{-4},10^2]$ (with spacing $0.5\,\mathrm{dex}$, or $0.25\,\mathrm{dex}$ for $\rho\in[10^{-1},10^1]$, where the variation is fastest). Here, the upper bounds are set by computational restrictions; models with large $\rho$ and $c_{\rm rel}$ feature fast subhorizon oscillations which are difficult to resolve numerically (and do not source interesting phenomenology). We use the following mass points:
\beq
    \sqrt{m^2/H^2-9/4}\in\{1.49i,1.4i,1.25i,1.0i,0.5i,0.01,0.25,0.5,\cdots,5.0\},
\eeq
covering both the complementary and principal series. Since the computational cost grows exponentially with the triangle squeezing, we omit configurations with $x< 0.01$; these are filled in using the known analytic squeezed limits \eqref{eq: strong-squeezed} (with $S_\zeta \sim x+\mathcal{O}(x^3)$ for the non-oscillatory pieces), fit to $x\in[0.01,0.1]$. 

Below, we will find that many shapes templates have detectable amplitudes only for $\rho/H\gtrsim 100$. To capture this regime, we supplement the above shapes with templates computed using the non-local single-field Lagrangian of \eqref{eq: lagrangian-cosmoflow-nonlocal}, which provides an accurate model for the correlators at large effective mass. These are computed similarly using \textsc{CosmoFlow}; following initial testing, we adopt absolute and relative tolerances of $10^{-80}$ and $10^{-10}$. To ensure that the two regimes smoothly connect, we compute the non-local results for $\rho\in[1,10^3]$, using the same spacing as before. In total, we compute over $40\,000$ templates (across all interactions and quadratic parameters, each for $145$ ($x,y$)-points), with a total of more than four million \textsc{CosmoFlow} calls. This has a total computational cost of around $10^5$ CPU-hours.

Fig.\,\ref{fig: strong-shape-comparison} compares the single exchange templates computed using the \textsc{CosmoFlow} and bootstrap pipelines as a function of $\rho, c_\pi$ and $m$ (see \citep{cosmoflow2} for further discussion). For $\rho\ll H$ the agreement is excellent across all masses and sound-speeds, validating both approaches.\footnote{We have additionally verified that the \textsc{CosmoFlow} predictions for $\dot\pi^2\sigma$ closely match the bootstrap results for $\rho/H\ll1$, and those for $(\partial_\mu\pi)^2\dot\pi$ and $\dot\pi^3$ agree with the analytic self-interaction shapes \eqref{eq: no-mixing-templates}.} In weak mixing (assumed in the bootstrap method), the template amplitude scales as $\rho^2$; from the figure, we find that this assumption breaks down for $\rho\gtrsim m$, and the amplitudes grow much more slowly.\footnote{This is partly due to the renormalization of $f_\pi$ required to offset power spectrum enhancements.} Moreover, both the frequency and phase of the squeezed-limit oscillations changes as $\rho$ increases (cf.\,\ref{eq: strong-squeezed}), with the oscillation amplitude decaying exponentially with $m_{\rm eff}=\sqrt{m^2+\rho^2}$. The significant differences between the bootstrap and \textsc{CosmoFlow} results at modest $\rho/H$ indicate that (a) weak mixing approaches are appropriate only for $\rho\ll H$ and (b) exchange signatures at large $\rho$ are considerably suppressed.

Thus far, we have considered only the $(\partial_\mu\pi)^2\sigma$ interaction. In Fig.\,\ref{fig: strong-all-shapes}, we consider the full set of tree-level bispectra and their phenomenology as a function of $\rho$ and $c_{\rm rel}$ for a particular choice of exchange mass. At small $\rho/H$, the shape is dominated by single exchange diagrams and self-interactions, whilst higher-order interactions become relevant as $\rho$ increases, though eventually plateau. For $\rho/H\sim 1$, we find non-trivial squeezed-limit oscillations in all channels: these have the same frequencies (set by $m_{\rm eff}$) but differing phases and offsets. Notably, we find oscillations even in the $(\partial_\mu\pi)^2\dot\pi$ and $\dot\pi^3$ diagrams due to the large mixing between $\pi$ and $\sigma$. By $\rho=100H$, the oscillatory contributions to all channels have decayed away (except for small residual noise) and all shapes approach the single-field self-interaction forms, matching the conclusions above. In \S\ref{sec: results}, we will explore all of the above templates as a function of $m$ and $\rho$, in order to search for novel scalar exchange signatures.

\subsection{Observational Data}\label{subsec: data}
\noindent In this work, our dataset is the bispectrum shape function, $\widehat{S}_i \equiv \widehat{S}(x_i,y_i)$, inferred from observations of the CMB temperature and polarization anisotropies.\footnote{Publicly available at \href{https://github.com/oliverphilcox/Binned-Bispectrum-Shape}{github.com/oliverphilcox/Binned-Bispectrum-Shape}.} Specifically, we use the measurements from \citep{Philcox:2026njr}, which were obtained by applying quasi-optimal estimators for the scale-invariant bispectrum to component-separated maps from \textit{Planck} PR4 \citep{Planck:2020olo} (using the \textsc{sevem} component-separation pipeline \citep{Planck:2018yye}). Most inflationary studies work with the full scale-dependent bispectrum, $B_\zeta$, or the unprocessed anisotropy maps \citep{Komatsu:2003iq,Fergusson:2009nv,Planck:2019kim}; directly using shape function allows us to trivially compare data and theory, facilitating fast sampling of the high-dimensional likelihood. 

The measurements use $171$ logarithmically-spaced bins in $(x,y)$-space (obtained via an intermediate set of $5000$ bins in $(k_1,k_2,k_3)$-space), with $10^{-3}\leq x\leq 1$ and $0.5\leq y\leq 1$, including squeezed, folded, and equilateral regimes. These were computed using the \textsc{PolySpec} code \citep{Philcox:2023uwe,Philcox:2023psd,Philcox:2026njr}, and are accompanied by an estimate of the $(171\times 171)$ covariance matrix, $\mathbb{C}_{ij} \equiv \mathrm{Cov}(\widehat{S}_i,\widehat{S}_j)$, itself derived from high-resolution \textsc{ffp10/npipe} simulations. A variety of observational effects are included in the estimator, including the beam, mask, and noise; this allows the output shapes to be immediately compared to theoretical models.

\subsection{Likelihood \& Priors}\label{subsec: priors}
\noindent Starting from the grid of inflationary templates described in \S\ref{subsec: templates}, we build combined models for the weakly- and strongly-mixed bispectrum using \eqref{eq: weak-exchange} and \eqref{eq: strong-exchange} respectively. To smoothly account for the dependence on quadratic parameters, we linearly interpolate across $\log_{10}\rho$ and $\log_{10}c_{\rm rel}$, after whitening the templates to remove the leading scalings.\footnote{We do not interpolate across $(x,y)$ since the same parameter points are used for both the data and theory.} The net result is a binned theory model $S_i \equiv S(x_i,y_i)$, depending on all parameters of interest. We write the full model using the \textsc{jax} language \citep{jax2018github}, facilitating fast model evaluation (in milliseconds) and analytic gradients. 

Assuming Gaussian noise (valid under the central limit theorem), the likelihood of the quadratic and cubic couplings, $\Theta_2$ and $\Theta_3$, given the observed shape, $\widehat{S}_i$, is
\beq\label{eq: loglkl}
    \log\mathcal{L}(\Theta_2,\Theta_2) = -\frac{1}{2}\sum_{ij}\left(S_i(\Theta_2,\Theta_3) - \widehat{S}_i\right)\mathbb{C}^{-1}_{ij}\left(S_j(\Theta_2,\Theta_3) - \widehat{S}_j\right) - \frac{1}{2}\log|\mathbb{C}| + \text{const.}
\eeq
Since the model is linear in $\Theta_3$, we can simplify the analysis using analytic marginalization. Writing $\mathcal{L}_{\rm marg}(\Theta_2) = \int d\Theta_3 \mathcal{L}(\Theta_2,\Theta_3)p(\Theta_3)$ for Gaussian prior $p(\Theta_3)$ with mean $\overline\Theta_3$ and covariance $\mathbb{P}$ and integrating leads to the marginalized likelihood:
\beq\label{eq: loglkl-marg}
    \log\mathcal{L}_{\rm marg}(\Theta_2) = -\frac{1}{2}\sum_{ij}\left(S_i(\Theta_2,\overline\Theta_3) - \widehat{S}_i\right)\mathbb{C}_{{\rm marg},ij}^{-1}(\Theta_2)\left(S_j(\Theta_2,\overline\Theta_3) - \widehat{S}_j\right) - \frac{1}{2}\log|\mathbb{C}_{{\rm marg}}(\Theta_2)| + \text{const.},
\eeq
where $\mathbb{C}_{\rm marg}(\Theta_2) = \mathbb{C} + (\nabla_{\Theta_3}S)\cdot\mathbb{P}\cdot(\nabla_{\Theta_3}S)^\dagger$.\footnote{An analogous approach can be used to analytically marginalize over parameters with flat priors.} A similar expression can be used to compute the \textit{maximum a posteriori} (MAP) point, dropping the $\log|\mathbb{C}_{\rm marg}(\Theta_2)|$ marginalization penalty. This significantly expedites sampling by reducing the effective dimension; moreover, one can draw posterior samples of $\Theta_3$ at negligible cost given the conditional distribution
\beq
    \Theta_3(\Theta_2) \sim \mathcal{N}\!\left(\overline{\Theta}_3 - \Sigma(\Theta_2)\,(\nabla_{\Theta_3}S)^\dagger,\,
    \Sigma(\Theta_2)\right) \quad \text{for} \quad
    \Sigma^{-1}(\Theta_2) = (\nabla_{\Theta_3}S)^\dagger\,\mathbb{C}^{-1}\,(\nabla_{\Theta_3}S)
    + \mathbb{P}^{-1}.
\eeq

For weak mixing, the model is specified by $\Theta_2=\{c_\pi,c_\sigma,\rho\}$ and $\Theta_3=\{\tilde c_3, \Delta\rho\}$, fixing the mass parameter. We adopt the following priors:
\beq\label{eq: weak-priors}
    &&\log_{10}c_\pi \sim \mathcal{U}(-2,0), \quad\log_{10}c_\sigma \sim \mathcal{U}(-2,0),\quad\tilde{c}_3(c_\pi^{-2}-1) \sim \mathcal{N}(0, (2\pi\Delta_\zeta^2)^{-1})\\\nonumber
    &&\rho^2/H^2 \sim \mathcal{U}(0,10^6), \quad \rho\Delta\rho/H^2 \sim \mathcal{U}(-10^6,10^6),
\eeq
where $\mathcal{U}$ and $\mathcal{N}$ represent uniform and Gaussian distributions. Here, the boundaries  on $c_\pi$ and $c_\sigma$ are fixed by the interpolator limits (\S\ref{subsec: templates}), with the width of $\tilde{c}_3(c_\pi^{-2}-1)$ set by the (approximate) unitarity bound (\S\ref{subsec: bounds}), which additionally require $c_\pi \gtrsim 0.01$. Since the weak mixing model is linear in $\rho^2$ and $\rho\Delta\rho$, we sample these parameters directly using wide priors. We do not impose perturbativity restrictions on $\rho$ and $\Delta\rho$ at this stage; later, we will use these to assess the viability of the weak mixing scenario.

For strong mixing, we have $\Theta_2=\{c_\pi,c_\sigma,\rho\}$ and $\Theta_3=\{\tilde c_3, \Delta\rho,\alpha,\mu\}$. For the quadratic parameters, we impose flat priors given the ranges used in \S\ref{subsec: templates} (recalling that $c_{\rm rel}\in[10^{-2},10]$):
\beq\label{eq: strong-priors}
    &&\log_{10}c_\pi \sim \mathcal{U}(-2,0), \quad\log_{10}c_\sigma \sim \mathcal{U}(-1,0),\quad \rho \sim \mathcal{U}(10^{-4},\rho_{\rm max}(\Theta_2)), \quad \tilde{c}_3(c_\pi^{-2}-1) \sim \mathcal{N}(0, (2\pi\Delta_\zeta^2)^{-1})\\\nonumber
    &&\Delta\rho \sim \mathcal{N}(0,\Delta\rho^2_{\rm max}(\Theta_2)), \quad \alpha \sim \mathcal{N}(0,\alpha^2_{\rm max}(\Theta_2)), \quad \mu \sim \mathcal{N}(0,\mu^2_{\rm max}(\Theta_2))
\eeq
where $X_{\rm max}(\Theta_2)$ are the (approximate) theoretical bounds given in \S\ref{subsec: bounds}.\footnote{Since the strong mixing bounds in \S\ref{subsec: bounds} require $\rho\gtrsim H$, we use the larger of \eqref{eq: bounds-weak}\,\&\,\eqref{eq: bounds-weak} for each $\rho$. We additionally truncate $\rho_{\rm max}$ to $10^{3}$.} Note that we impose a uniform prior on $\rho$ since it enters the model non-linearly (in contrast to the uniform $\rho^2$ prior used for weak mixing). An alternative choice would be to impose a uniform prior on $\log_{10}\rho$; this is explored in Appendix \ref{app: alternative-prior}. We additionally place informative Gaussian priors on all cubic parameters given the approximate unitarity bounds: this ensures that the regimes probed by the model are theoretically well-motivated.

To perform the analysis, we sample \eqref{eq: loglkl-marg} using both nested sampling (via the \textsc{dynesty} code \citep{Speagle:2019ivv}) and Hamiltonian Monte Carlo (HMC, via \textsc{numpyro} \citep{phan2019composable}), reconstructing linear parameters from the chains as described above. To quantify the preference for a given model over the null hypothesis, we will compute both $\Delta\chi^2 = \chi^2_{\rm MAP}-\chi^2_{\rm null}$ and the Bayesian evidence ratio, $\log B = \log(\mathcal{Z}/\mathcal{Z}_{\rm null})$. For the former, we use the differential evolution algorithm, supplemented with the Limited-memory Broyden–Fletcher–Goldfarb–Shanno (L-BFGS) approach (with multiple start-points across the $\Theta_2$-space), whilst the latter is obtained using the nested sampler. We will additionally compute marginalized upper bounds on various parameters, $X_{95}$: these are obtained from the output chains via $\int_{X_{95}}^\infty  dX\,\mathcal{P}(X)=0.05$.

\section{Results}\label{sec: results}
\noindent We now present the \textit{Planck} constraints on the multi-field inflationary lagrangian. We will start from the simple single-field limit ($\rho=0$), before introducing a weakly-mixed field ($\rho\ll H$) and finally extending to the full scenario with arbitrary $\rho/H$. As discussed in \citep{cosmoflow2}, the weak mixing results are broadly incompatible with the theoretical restrictions in \S\ref{subsec: bounds}; however, their inclusion serves as a useful warm-up exercise for the full scenario and facilitates comparison with other studies, which (either implicitly or explicitly) assume $\rho\lesssim H$.

At fixed mass, the complete system is a function of seven parameters: $\{c_\pi,c_\sigma, \rho, \tilde{c}_3,\Delta\rho,\alpha,\mu\}$ (Tab.\,\ref{tab: parameters}). To simplify the treatment, we will consider a number of limiting regimes in addition to the full ensemble: (1) analyzing each physical signature independently, e.g., constraining $\alpha$ with $\mu=\Delta\rho=\tilde{c}_3=0$; (2) ignoring self-interactions by setting $c_\pi=1$ (which removes the enhancement of spatial gradients); (3) fixing the scalar sound-speed to $c_\sigma=1$ (matching most previous works). 

\subsection{No Mixing}\label{subsec: results-no}

\begin{figure}[!t]
    \centering
    \includegraphics[width=0.4\linewidth]{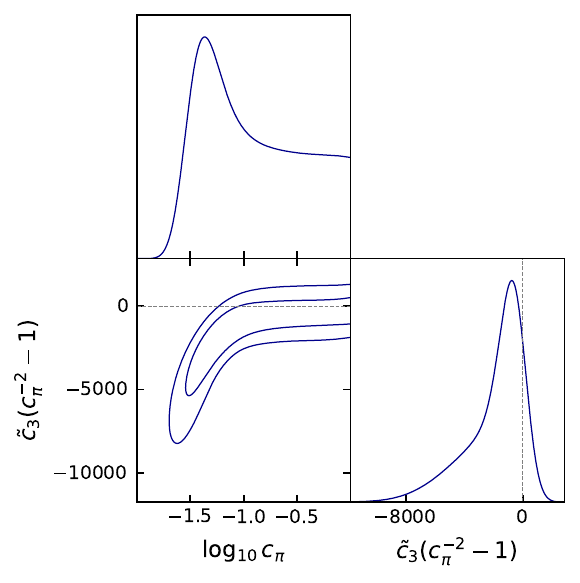}
    \caption{\textbf{No Mixing Results}: Posterior on the single-field parameters $c_\pi$ and $\tilde{c}_3$ from \textit{Planck} CMB data. We find no evidence for self-interactions, with $\Delta\chi^2=-2.0$ and the (log) Bayes factor $\log\mathcal{Z}/\mathcal{Z}_{\rm null} = -2.2$. Using the setup described in the text, we find the $95\%$ constraints $c_\pi > 0.029$ and $|\tilde{c}_3(c_\pi^{-2}-1)|<6080$, which are slightly wider than those of \citep{Planck:2019kim} due to differing prior choices.}
    \label{fig: results-no-mixing}
\end{figure}

\noindent First, we analyze the single-field inflationary model, setting $\rho=0$. The resulting bispectra depend on two parameters: $c_\pi$ and $\tilde{c}_3$, which we sample according to \eqref{eq: weak-priors}, including a mild perturbativity prior on $\tilde{c}_3(c_\pi^{-2}-1)$. As shown in Fig.\,\ref{fig: results-no-mixing}, we report no detection of primordial non-Gaussianity, with the $95\%$ bounds $c_\pi > 0.029$ and $|\tilde{c}_3(c_\pi^{-2}-1)|<6080$ (with the $1\sigma$ limit $-1800<\tilde{c}_3(c_\pi^{-2}-1)<2100$). These constraints are similar to the frequentist bounds obtained in \citep{Planck:2019kim} through analyses of the $f_{\rm NL}^{\rm eq}$ and $f_{\rm NL}^{\rm orth}$ parameters (see also \citep{Senatore:2009gt,2014A&A...571A..24P,Planck:2015zfm,Philcox:2025wts}), though our $c_\pi$ bound is somewhat weaker since we (a) vary $\log_{10}c_\pi$ instead of $c_\pi^{-2}$ and (b) impose a weak unitarity prior on $\tilde{c}_3(c_\pi^{-2}-1)$, which disfavors the low-$c_\pi$ tail.\footnote{Imposing a uniform prior on $c_\pi^{-2}$ and inflating the prior on $\tilde{c}_3(c_\pi^{-2}-1)$, we find the 95\% bounds $c_\pi> 0.022$ and $|\tilde{c}_3(c_\pi^{-2}-1)|<9170$, in good agreement with the Bayesian analysis of \citep{Philcox:2025wts}.} Notably, our bounds are consistent with the theoretical limits of \S\ref{subsec: bounds}, implying that current data can meaningfully constrain the single-field scenario.

To quantify the preference of the data for the single-field EFT model, we compute both the best-fit $\chi^2$ and the Bayes factor relative to the fiducial model: $c_\pi=1,\,\tilde{c}_3=0$ (corresponding to $S_\zeta=0$). At the MAP, $\chi^2$ improves by $2.0$; the addition of two degrees of freedom leads to the Bayesian evidence reducing, however, with $\log\mathcal{Z}/\log\mathcal{Z}_{\rm null}=-2.2$. As found in previous works \citep[e.g.,][]{Planck:2019kim}, this indicates that the single-field model is not observationally favored.

\subsection{Weak Mixing}\label{subsec: results-weak}

\begin{figure}
    \centering
    \includegraphics[width=0.8\linewidth]{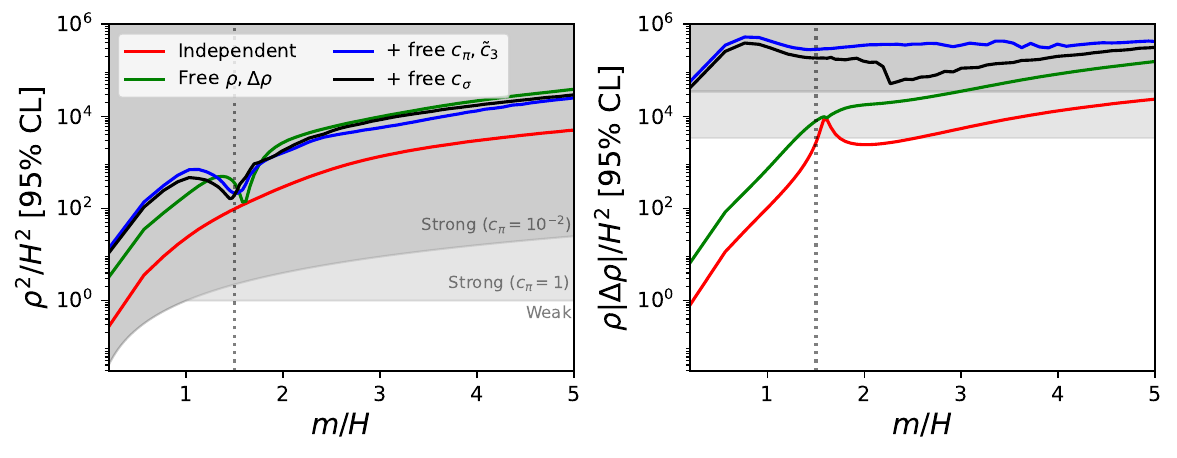}
    \includegraphics[width=0.8\linewidth]{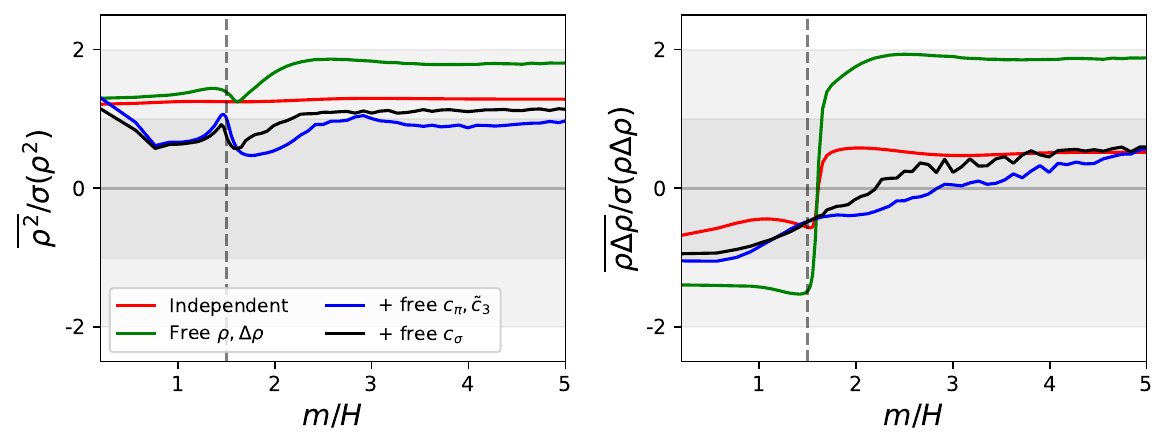}
    \includegraphics[width=0.8\linewidth]{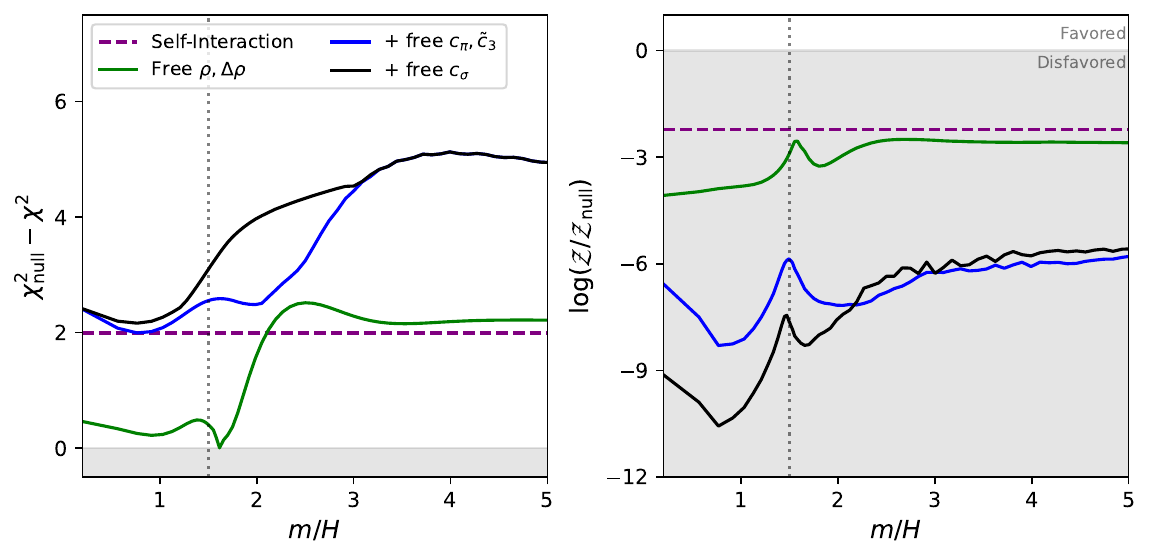}
    \caption{\textbf{Weak Mixing Results}: 
    \textit{Upper panels}: 95\% bounds on the single exchange amplitudes $\rho^2$ and $\rho\Delta\rho$, corresponding to the $(\partial_\mu\pi)^2\sigma$ and $\dot\pi^2\sigma$ vertices, with templates computed using bootstrap methods, assuming the weakly-mixed regime ($\rho\lesssim H$). We plot four sets of constraints as a function of scalar mass, $m$: analyzing each signature independently (red); jointly analyzing $(\partial_\mu\pi)^2\sigma$ and $\dot\pi^2\sigma$ at fixed $c_\pi=c_\sigma=1$ and $\tilde{c}_3=0$ (green); additionally including self-interactions with free $c_\pi$ and $\tilde{c}_3$ (blue); additionally varying the exchange speed $c_\sigma$ (black). The light (dark) gray panels show the perturbativity bounds for $c_\pi=1$ ($c_\pi = 0.01$), above which the weak mixing assumption breaks down. Except for boost-breaking interactions with $m\lesssim 3H/2$, the constraints are incompatible with the weak mixing assumptions \citep[cf.][]{cosmoflow2}; this motivates the more general analyses of Fig.\,\ref{fig: results-strong-mixing}.
    \textit{Middle panels}: Signal-to-noise ratio for the $(\partial_\mu\pi)^2\sigma$ and $\dot\pi^2\sigma$ interactions. 
    We find no detection above $2\sigma$, regardless of the analysis type. 
    \textit{Bottom panels}: Preference for the weak mixing model compared to the null hypothesis as indicated by the $\chi^2$ improvement (left) and Bayes factor (right). The purple line gives the results for the single-field theory (with $\rho=0$). Since $\mathcal{Z}<\mathcal{Z}_{\rm null}$ everywhere (and further $\mathcal{Z}<\mathcal{Z}_{\rm single-field}$), we report no detection of the weakly-mixed model.}
    \label{fig: results-weak-mixing}
\end{figure}

\noindent Next, we consider the weakly-mixed scenario, using the theoretical templates computed with the cosmological bootstrap (\S\ref{subsec: templates}). To begin, we present the bounds on the coupling amplitudes $\rho^2$ and $\rho\Delta\rho$, analyzed independently with $c_\pi=c_\sigma=1$ and $\tilde{c}_3=0$. As shown in Fig.\,\ref{fig: results-weak-mixing}, the constraints are a strong function of the exchange mass, with $\rho_{\rm 95}$ ranging from $0.5H$ at $m\approx 0.2H$ to $70H$ at $m\approx5H$. This is expected: the signal-to-noise for $m\lesssim3H/2$ is dominated by the squeezed-limit divergence (with $S_\zeta\to x^{-1}$ as $m\to 0$, cf.\,a\ref{eq: weak-squeezed}), while the shapes with $m\gg 3H/2$ are dominated by the self-interaction-like equilateral contributions, whose amplitudes scale as $(m^2/H^2-9/4)^{-1}$ (\S\ref{subsec: pheno}). Jointly analyzing the two single exchange templates weakens the bounds by around a factor of ten, confirming that the two signatures are highly correlated. Adding the self-interaction signatures further degrades the $\rho\Delta\rho/H^2$ constraints at low $m/H$, indicating that the data is sensitive to only one of the two exchange diagrams.\footnote{This occurs since both shapes are dominated by the same squeezed limit divergence, and is the reason why one can use a single amplitude, $f_{\rm NL}^{\rm local}$, to parametrize all exchange diagrams in the $m\to 0$ limit.} At large $m/H$, marginalizing over self-interactions does not appreciably alter the $\rho^2$ constraints. While this may seem surprising since (a) the exchange shapes become featureless for $m\gg H$ and thus well-correlated with self-interaction signatures and (b) the $(\partial_\mu\pi)^2\sigma$ signals are enhanced for small $c_\pi$ (\S\ref{subsec: pheno}), it is rationalized by noting that marginalization over $\rho\Delta\rho$ already decorrelates much of the featureless contribution and most of the posterior mass lies around $c_\pi \approx 1$ where the single-field terms are small. Finally, we note that marginalizing over $c_\sigma$ only minorly changes the constraints; this follows expectations since $c_\sigma$ primarily modulates the phase of the oscillations rather than the amplitudes. 

\begin{figure}
    \centering
    \includegraphics[width=0.8\linewidth]{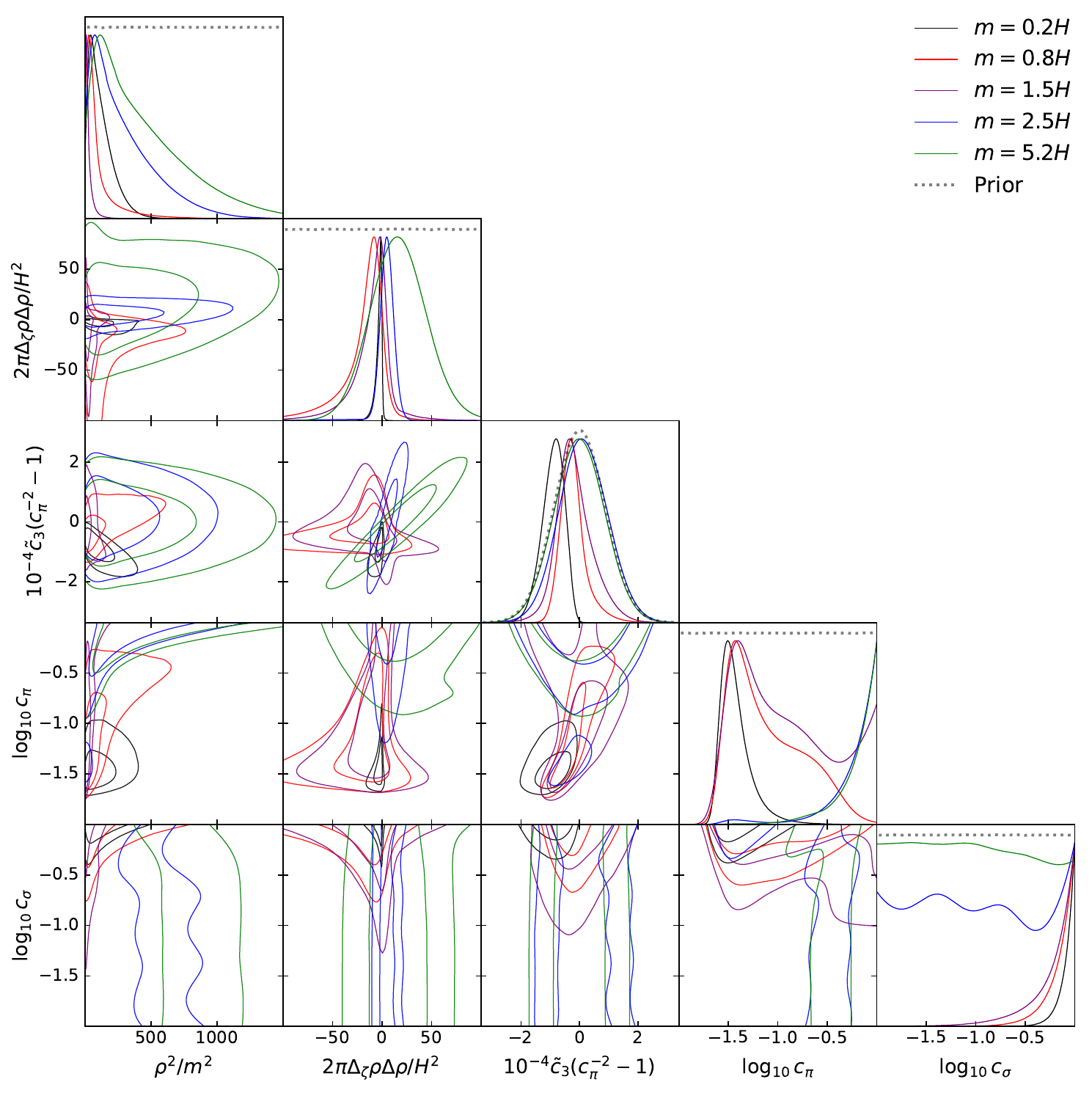}
    \caption{\textbf{Joint Constraints on the Weak Mixing Collider}: Posterior on the quadratic and cubic parameters describing the two-field inflationary Lagrangian, computed using bootstrap techniques. We show results for five choices of mass, across both the complementary and principal series. In all cases, the posteriors display highly non-trivial correlations, as seen, for example, in the rotating degeneracy direction between $\rho\Delta\rho$ and $\tilde{c}_3(c_\pi^{-2}-1)$. For $m\gtrsim H$, the data prefer $c_\pi\approx 1$ (leading to vanishing self-interaction signatures); at lower masses, $c_\pi \approx 10^{-1.5}$ is favored, though this does not translate into a strong preference for non-Gaussianity (Fig.\,\ref{fig: results-weak-mixing}). Variations of the scalar sound-speed, $c_\sigma$, shift the phase of oscillations in the exchange bispectra; this manifests as wiggles in the (well-converged) contours. We caution that these results do not impose perturbativity bounds on $\rho$ and $\Delta\rho$ -- as shown in Fig.\,\ref{fig: results-weak-mixing}, most detectable scenarios require $\rho\gtrsim H$, which must be modeled using alternative techniques (cf.\,Fig.\,\ref{fig: results-strong-mixing}).}
    \label{fig: corner-weak-mixing}
\end{figure}

The second panel of Fig.\,\ref{fig: results-weak-mixing} shows the signal-to-noise ratio on the two exchange amplitudes. Across all masses and analysis variants, we find a maximum detection significance of $1.9\sigma$ at $m=2.4H$, giving no evidence for novel primordial physics. At large $m$, the signal-to-noise plateaus: in this limit, the model is well-described by single-field effective field theory (as one can integrate-out $\sigma$), and thus the well-known equilateral and orthogonal templates \citep{Creminelli:2005hu,Senatore:2009gt}. To assess the correlation between the various parameters, we plot their joint distribution in Fig.\,\ref{fig: corner-weak-mixing} for a representative set of masses. Due to the significant correlations between templates as well as the non-linear behavior of the shapes themselves, the full posterior is highly complex with a form depending sensitively on $m$. A few general features are of note: (a) the $\rho\Delta\rho$ and $\tilde{c}_3$ parameters are highly correlated even at low mass (sourcing the degradation seen in Fig.\,\ref{fig: results-weak-mixing}); (b) constraints on $\rho^2$ and $\rho\Delta\rho$ are a strong function of $c_\pi$, and, for $m\gtrsim H$, tighten strongly when $c_\pi\ll 1$ due to the slow-collider enhancements \citep{Jazayeri:2022kjy}; (c) low-mass templates have a preference for $c_\pi\ll 1$ (d) $c_\sigma$ modulates the squeezed-limit oscillations, giving wiggles in the output contours. 

Given these results, one can ask whether the \textit{Planck} data displays a net preference for the weakly-mixed multi-field scenario. Ignoring self-interactions (thus fixing $c_\pi=1$), $\chi^2$ reduces by at most $2.5$ (at $m=2.5H$); this changes to $5.2$ (at $m=4H$) when jointly sampling self-interactions and exchange contributions. At large $m/H$, the $\chi^2$ values stabilize (with a preference for $c_\sigma=1$) since the templates asymptote to featureless non-local forms (cf.\,Fig.\,\ref{fig: weak-shapes}). The reduction in $\chi^2$ is not reflected in the Bayesian evidence ratios: $\mathcal{Z}<\mathcal{Z}_{\rm null}$ for all models considered, indicating that the scenario is not supported by data.\footnote{An important limitation of the Bayes factor is its dependence on the (often \textit{ad hoc}) prior widths. To circumvent this we adjust the priors on $\rho^2$ and $\rho\Delta\rho$ used in the $\log \mathcal{Z}$ analyses to have mass-dependent width $\pm5\sigma$ rather than the broad priors of \eqref{eq: weak-priors}. This increases $\log\mathcal{Z}$ (avoiding the penalty from integrating up to $\rho^2\leq 10^6$ when the low-mass data constrain $\rho^2<1$ for example). As alternatives, we also compute the Akaike and Bayesian information criteria: $\mathrm{AIC} = \chi^2+2k, \,\mathrm{BIC} = \chi^2+k\log n$ (for $k$ parameters and $n$ data-points) \citep{Akaike:1974vps,Schwarz:1978tpv} In both cases, we find no preference for the model compared to the null hypothesis, with $\Delta\mathrm{AIC}>1$ and $\Delta\mathrm{BIC}>5$.} Moreover, the Bayesian evidences are below those of the self-interaction scenario in \S\ref{subsec: results-no}, further disfavoring the scenario.

Lastly, we compare the model constraints to the bounds given in \S\ref{subsec: bounds}. As seen in Fig.\,\ref{fig: results-weak-mixing}, the \textit{Planck} constraints on $\rho^2$ are far weaker than the weakly-mixed perturbativity bound for all $m\gtrsim 0.2H$, even in the most optimistic scenario (fixing $c_\pi=c_\sigma=1$ and $\tilde{c}_3=\Delta\rho=0$). For the boost-breaking shape, the bounds on $\Delta\rho$ are competitive only for $m\lesssim 3H/2$ (the complementary series), in agreement with \citep{Pinol:2023oux,cosmoflow2}.\footnote{When marginalizing over $c_\pi$ and $\tilde{c}_3$, the shape appears unmeasurable; this is due to the large correlations between the $(\partial_\mu\pi)^2\sigma$ and $\dot\pi^2\sigma$ templates and can be avoided by an appropriate orthogonalization.} At low $c_\pi$, the perturbativity bounds on $\Delta\rho$ weaken; however, the (necessary) marginalization over self-interactions prohibits detectably-large signals. This is an important conclusion: within the weak-mixing regime ($\rho\lesssim H$), current data cannot place meaningful constraints on scalar exchange except at low masses, where a single shape dominates (in the `quasi-single-field' regime) \citep{Chen:2009we,Chen:2009zp,Noumi:2012vr}.\footnote{While we use a somewhat different setup to \citep{cosmoflow2}, marginalizing over self-interactions rather than projecting out equilateral and orthogonal templates, our conclusions are similar.} Notably, this is an \textit{analysis} limitation not a \textit{physics} limitation: the restrictive bounds shown in Fig.\,\ref{fig: results-weak-mixing} arise since we choose to treat the $\pi-\sigma$ coupling perturbatively. This motivates the general scenario discussed in the next section, which allows for arbitrary $\rho/H$.

\subsection{Strong Mixing}
\noindent We now turn to the main results of this work: constraints on the collider physics with arbitrary $\rho/H$. As detailed in \S\ref{subsec: strong-theory}, the exchange signatures depend on four coupling amplitudes, $\{\rho,\Delta\rho,\alpha,\mu\}$, as well as two sound-speeds, a cubic self-interaction coefficient, and the scalar mass (cf.\,Tab.\,\ref{tab: parameters}). 
A crucial difference to the previous analysis is that we treat the quadratic mixing non-perturbatively using \textsc{CosmoFlow}; this allows us to probe regimes where the classic division into amplitude and $\rho$-independent shape breaks down. As we will see below (see also \citep{cosmoflow2}), this regime is both accessible to current observations and not \textit{a priori} ruled out by theoretical constraints.

\begin{figure}
    \centering
    \includegraphics[width=\linewidth]{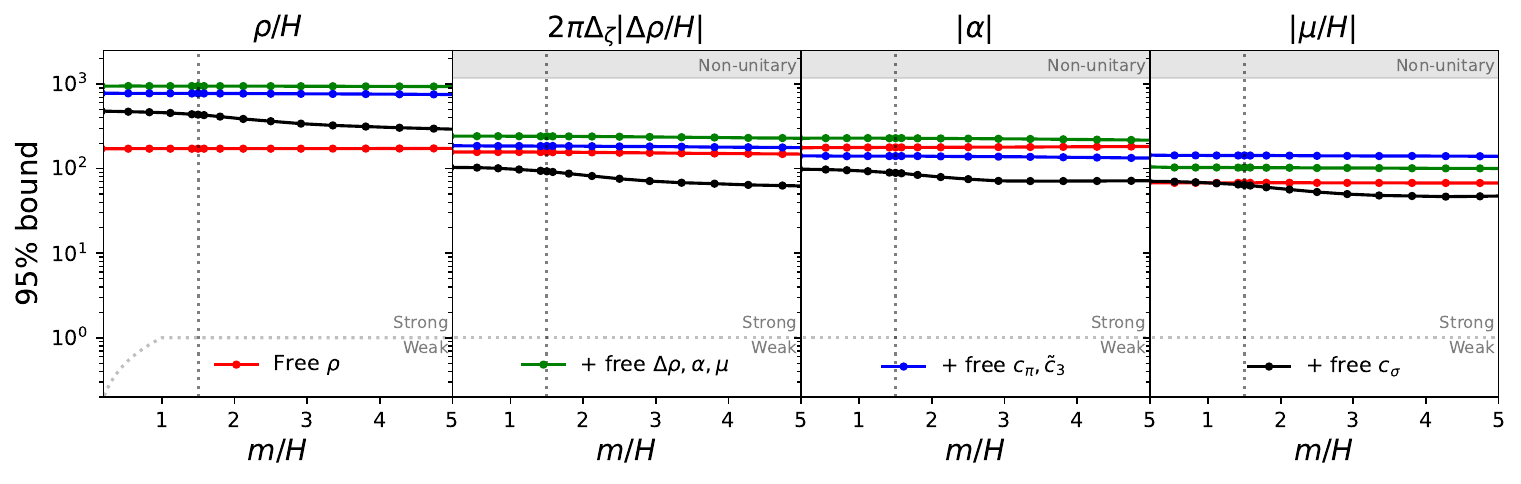}
    \includegraphics[width=\linewidth]{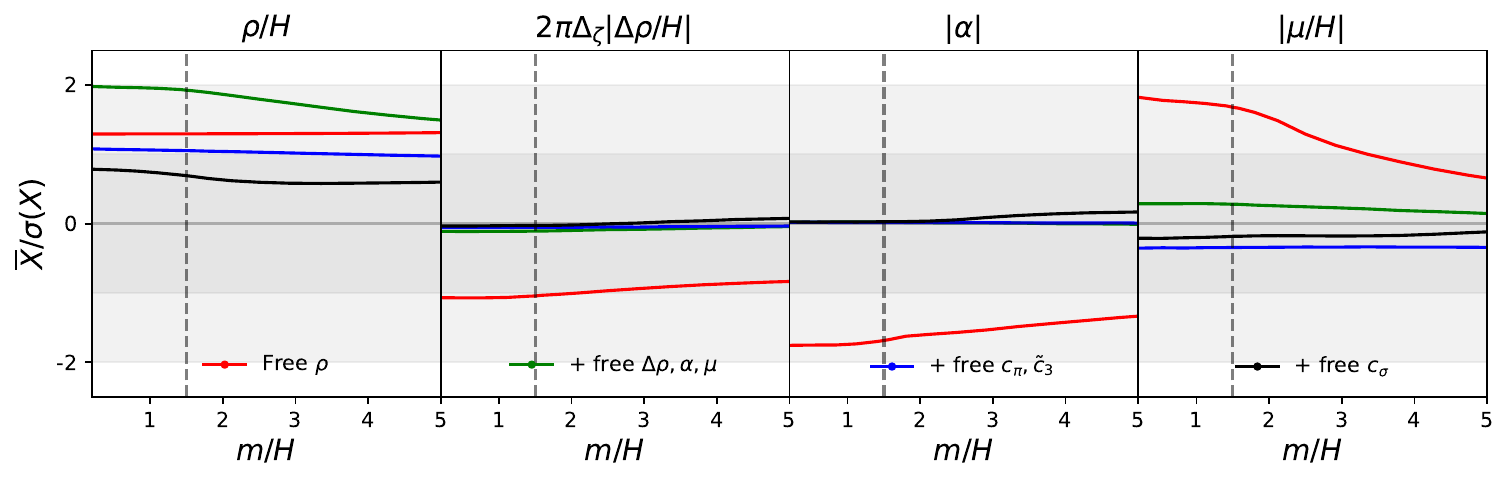}
    \includegraphics[width=0.8\linewidth]{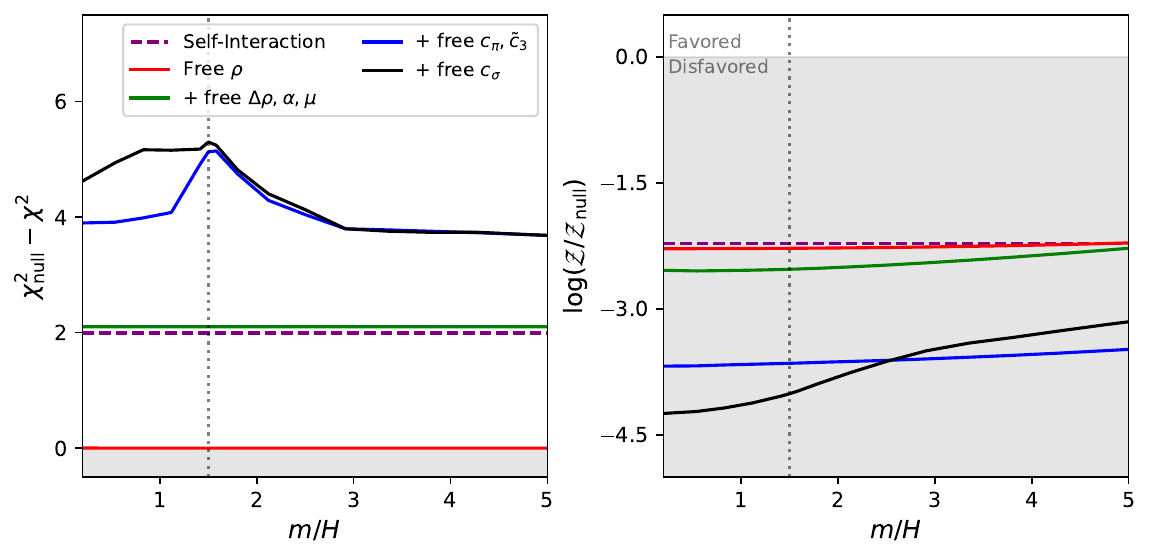}
    \caption{\textbf{Strong Mixing Results}: \textit{Upper panels}: 95\% constraints on the multi-field inflationary Lagrangian obtained from \textit{Planck}. We show constraints on the single-, double-, and triple-exchange amplitudes as a function of mass for four scenarios: marginalized over $\rho$ (red); adding all cubic couplings (green); adding self-interactions (blue); adding a free scalar sound-speed (black). The gray band shows the regions inconsistent with unitarity (\ref{eq: bounds-strong}, setting $c_\pi=c_\sigma=1$), while the dotted line indicates the weakly-mixed regime (assumed in Fig.\,\ref{fig: results-weak-mixing}). Our bounds are significantly tighter than the theoretical limits, but require strong mixing. The results are sensitive to the prior on $\rho$; constraints using a logarithmic prior are shown in Fig.\,\ref{fig: results-strong-mixing-log}. \textit{Middle panels}: Signal-to-noise ratio for the four operators, marginalized over $\rho$. Across all masses and analyses, we find no strong detections, with a maximum signal-to-noise of $1.83\sigma$ in independent analyses (for $\mu$ at $m=0.2H$) or $1.98\sigma$ in joint analyses ($\rho$ at $m=0.2H$).
    \textit{Bottom panels}: Preference for the cosmological collider. Though $\chi^2$ improves by up to $5.3$ units with respect to the null hypothesis ($\rho=0$, $c_\pi=1$) or $3.3$ units with respect to the single-field limit (dashed purple), we find no evidence for the collider scenario, with $\mathcal{Z}<\mathcal{Z}_{\rm null}$ in all cases.} 
    \label{fig: results-strong-mixing}
\end{figure}



First, we consider the constraints on the four coupling amplitudes as a function of exchange mass, $m$. In the simplest scenario of unit $c_\pi=c_\sigma$ with vanishing cubic interactions ($\tilde{c}_3=\Delta\rho=\alpha=\mu=0$), we find the 95\% bound $\rho\lesssim 170H$, approximately independent of mass.\footnote{This is sensitive to the choice of prior; as shown in Appendix \ref{app: alternative-prior}, switching to a logarithmic prior on $\rho/H$ leads to $\rho\lesssim 50H$.} Though far outside the weak mixing regime (which requires $\rho\lesssim H$), this is fully consistent with the unitarity bounds given in \S\ref{subsec: bounds}, which limit $\rho\lesssim 5000$ at $c_\pi=1$. The independence of $\rho_{95}$ on mass may seem surprising since the shape function has a strong squeezed-limit divergence for $m\ll H$; this arises since the induced single-exchange signal is too small to be detected for all $\rho$ within the weak mixing regime (as seen in the left panel of Fig.\,\ref{fig: results-weak-mixing}). To produce a detectably large signal, we require $\rho\gtrsim m$, which modifies the squeezed limit behavior (replacing $m$ by $m_{\rm eff}=\sqrt{m^2+\rho^2}$ following \ref{eq: strong-squeezed}), practically nulling the divergence and giving similar shapes for all $m$.

Also shown in Fig.\,\ref{fig: results-strong-mixing} are constraints on the cubic exchange amplitudes marginalized over $\rho$. For each parameter, the \textit{Planck} bound is several orders of magnitude below the perturbativity bound yet significantly outside the weak mixing regime \citep[cf.,][]{cosmoflow2}. As before, we find uniform constraints across all masses, with $2\pi\Delta_\zeta|\Delta\rho|\lesssim 150H$, $|\alpha|\lesssim 180$ and $|\mu|\lesssim H$. The constraints weaken somewhat when all cubic operators are varied simultaneously, as is expected from the similar shapes seen in Fig.\,\ref{fig: strong-all-shapes}. Freeing $c_\pi$ and $c_\sigma$ leads to somewhat tighter constraints, since many of the signals are boosted at low sound-speed, though this improvement is tempered by the accompanying marginalization over self-interaction signatures and the $\dot\sigma\partial_i\pi\partial_i\sigma$ vertex.

\begin{figure}
    \centering
    \includegraphics[width=\linewidth]{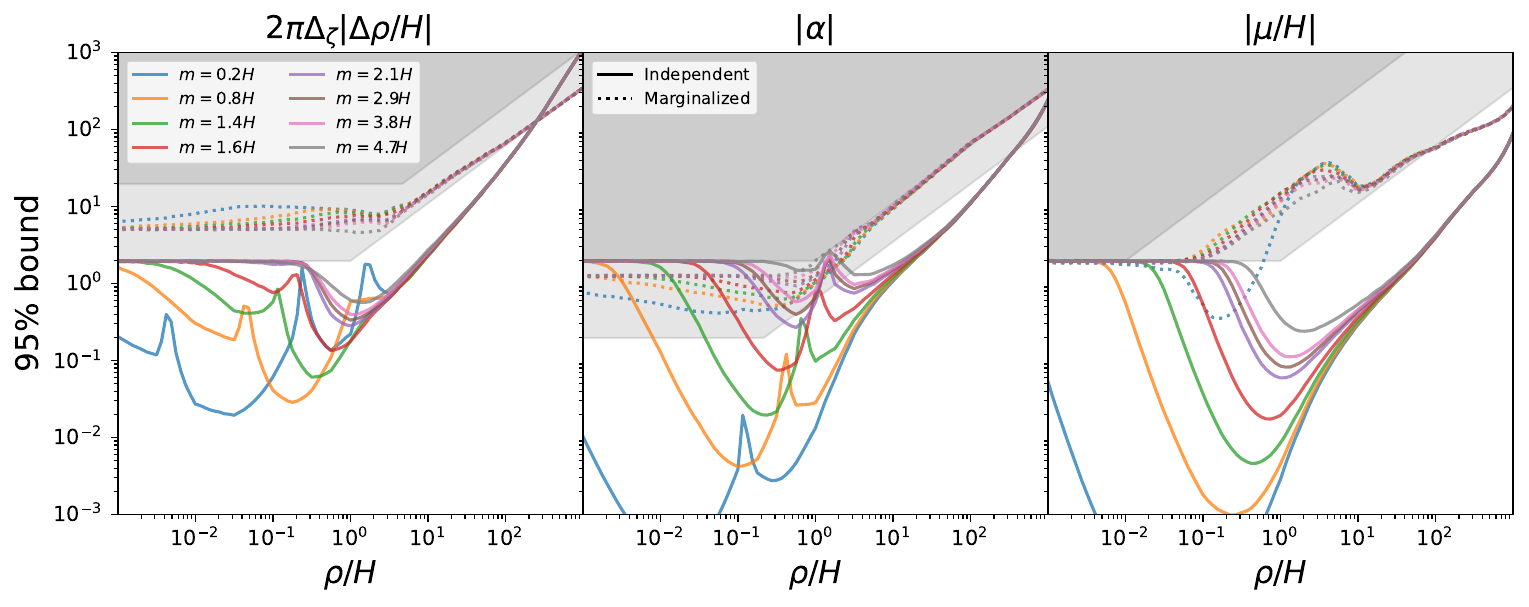}
    \caption{\textbf{Strong Mixing Results At Fixed $\rho$}: $95\%$ bounds on the cubic exchange amplitudes (cf.\,Fig.\,\ref{fig: cartoon}) as a function of quadratic mixing for various masses (indicated by colors). The grey regions show the (approximate) theoretical bound for $c_\pi\in\{0.01,1\}$ using \eqref{eq: bounds-strong}, which are used as Gaussian priors in the analysis. We show results for both a single cubic operator at $c_\pi=1$ (solid) and the full set of seven operators given in Tab.\,\ref{tab: parameters} (dotted). For $m\lesssim 3H/2$, the data places strong constraints on individual amplitudes in both weak ($\rho\lesssim H$) and strong ($\rho\gtrsim H$) scenarios, while for larger masses, the weak mixing regime is prior-dominated. Marginalization over the full seven-parameter system significantly weakens the bounds, though constraints are still competitive with respect to the theoretical limits (particularly for the triple-exchange amplitude).}
    \label{fig: results-strong-mixing-fixrho}
\end{figure}

Due to marginalization over $\rho$, constraints on the cubic parameters are more difficult to interpret than for the weakly-mixed scenario. To this end, we additionally perform analyses at fixed $\rho/H$, which are shown in Fig.\,\ref{fig: results-strong-mixing-fixrho}. The story is similar for all cubic exchange operators: as $m\to 0$, we obtain competitive constraints on the couplings for $\rho\lesssim H$, particularly for the triple-exchange diagram (with the $\Delta\rho$ behavior matching Fig.\,\ref{fig: results-weak-mixing}). For masses in the principal series ($m\geq 3H/2$), the constraints are broadly prior-dominated for $\rho\lesssim H$, reaffirming our earlier conclusion that weakly-mixed collider oscillations are difficult to probe observationally (in accordance with \citep{Pinol:2023oux,Kumar:2026dih,cosmoflow2}). At large $\rho/H$, we find strong constraints independent of $m$; these dominate the marginalized results shown in Fig.\,\ref{fig: results-strong-mixing}. Results from the joint analysis of all seven variables are broadly similar; small masses show some novel signatures at low $\rho/H$ (e.g., for double-exchange), but higher masses are prior-dominated at low-$\rho$. At large $\rho/H$, the dominant constraints arise from the triple-exchange channel, and remain within unitarity bounds even at $\rho=1000H$.

\begin{figure}[!t]
    \centering
    \includegraphics[width=\linewidth]{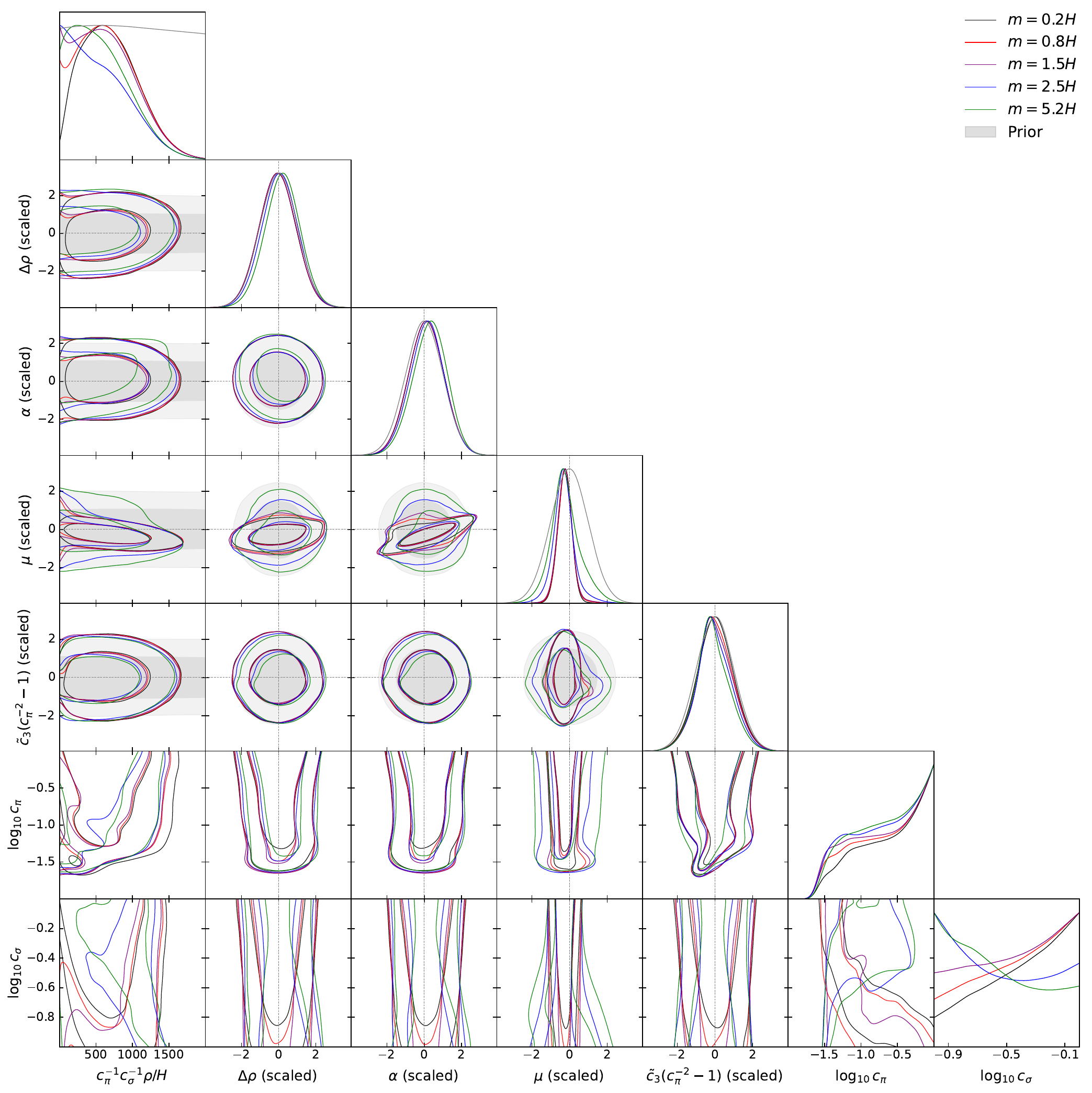}
    \caption{\textbf{Joint Constraints on the Strong Mixing Collider}: Posterior on the seven parameters describing the two-field inflationary Lagrangian obtained by analyzing the \textit{Planck} CMB bispectrum. We show results for five choices of mass (indicated by colors) as well as the prior (gray contours, omitted for the uniformly-sampled $\log_{10}c_\pi,\log_{10}c_\sigma$). For visualization, we normalize the linear parameters ($\Delta\rho,\alpha,\mu,\tilde{c}_3(c_\pi^{-2}-1)$) by the theoretical bounds (\eqref{eq: bounds-strong}), which are used as priors in the analysis. As in Fig.\,\ref{fig: results-strong-mixing}, we find no detection of non-Gaussianity, with many of the marginalized constraints reproducing the prior, though we find a complex set of degeneracies, with respect to $\rho$, $c_\pi$ and $c_\sigma$. The constraints show only mild dependence on mass, since the main phenomenology is set by the effective mass, $m_{\rm eff} = \sqrt{m^2+\rho^2}$. Analogous results using a logarithmic prior on $\rho$ are shown in Fig.\,\ref{fig: corner-strong-mixing-log}.}
    \label{fig: corner-strong-mixing}
\end{figure}

In Fig.\,\ref{fig: corner-strong-mixing} we show the joint constraints on the seven inflationary couplings. As in the weakly-mixed scenario (Fig.\,\ref{fig: corner-weak-mixing}), the posteriors have non-trivial degeneracy directions with respect to $\rho$, $c_\pi$ and $c_\sigma$ (arising since the underlying templates are set by $c_{\rm rel}=c_\pi/c_\sigma$). For $\Delta\rho, \alpha$ and $\tilde{c}_3(c_\pi^{-2}-1)$, the marginalized constraints reproduce the theory-motivated prior (though significant improvement is seen for the triple-exchange amplitude): as seen in Fig.\,\ref{fig: results-strong-mixing-fixrho}, this is a consequence of the integration over $\rho$. In agreement with Fig.\,\ref{fig: results-strong-mixing}, the posteriors show only mild dependence on mass; moreover, we find no clear evidence for a non-zero signal in any parameter. 

Finally, we consider the preference of \textit{Planck} for the multi-field scenario. The middle panel of Fig.\,\ref{fig: results-strong-mixing} shows the detection significance of each exchange amplitude, analyzed both independently and in tandem. Across all masses and configurations, we find a maximum signal-to-noise of $2.0\sigma$ (from the $\rho/H$ as $m\to 0$), which reduces to $1.8\sigma$ when analyzing each signature independently (from $\mu/H$ as $m\to 0$, marginalized over $\rho$). From this metric, we find no evidence for collider non-Gaussianity. At the maximum \textit{a posteriori} point, the $\chi^2$ reduces by $5.3$ relative to the null hypothesis, or $3.3$ with respect to the single-field theory, with no improvement found for the $\rho$-only scenario. Since the improvement in $\chi^2$ is accompanied by five additional free parameters, it does not indicate a preference for multi-field inflation according to either the Akaike or Bayesian information criteria. Moreover, the Bayesian evidences favor the null hypothesis for all masses and analysis variants. Overall, we conclude that the \textit{Planck} CMB data does not find evidence for the multi-field inflationary Lagrangian: this is a more meaningful statement than for the weak mixing scenario since our constraints are compatible with the theoretical restrictions.

\section{Conclusions}\label{sec: conclusions}
\noindent Multi-field inflation provides a wealth of observational signatures, including self-interactions and a tower of exchange diagrams. Though the most well-known signals are the cosmological collider oscillations arising in the squeezed limits, it is important to emphasize that these do not appear in isolation. To obtain a large scalar exchange signature that is consistent with theoretical bounds, we require either a small sound-speed or a large quadratic mixing: these distort the curvature bispectra by inducing large self-interaction signatures and strong modifications to the quadratic theory. To fully explore the multi-field scenario, we must \textit{jointly} constrain both the quadratic and cubic sectors, carefully modeling the induced bispectra and accounting for perturbativity constraints and non-linearly realized symmetries. 

In this work, we have performed a detailed analysis of the cosmological collider scenario, constraining the two-field effective Lagrangian using data from the \textit{Planck} satellite. This combines two important developments: (1) direct CMB measurements of the underlying inflationary shape function, $S_\zeta(x,y)$ \citep{Philcox:2026njr}; (2) non-perturbative methods for computing curvature bispectra from a given inflationary Lagrangian, allowing for arbitrary quadratic mixing \citep{Pinol:2023oux,Werth:2023pfl,Werth:2024aui}. Marrying the two, we have obtained precise predictions for the primordial shape functions (depending on four quadratic and four cubic parameters) that can compared against data in milliseconds, facilitating efficient Bayesian analyses that go far beyond the conventional template-matching approaches.

An important conclusion from both this work and our previous study \citep{cosmoflow2} is that large cosmological collider signatures cannot be generated in the weak mixing limit ($\rho\ll H$). This is clearly demonstrated in Fig.\,\ref{fig: results-weak-mixing}: for masses in the principal series (where the characteristic collider oscillations arise), current data are sensitive only to coupling amplitudes outside the ranges permitted by perturbativity. While this is not a new realization \citep[cf.][]{Pinol:2023oux,Kumar:2026dih}, it has been largely ignored by observational studies \citep{Cabass:2024wob,Suman:2025tpv,Suman:2025vuf,Sohn:2024xzd,Philcox:2026njr,Philcox:2025bbo,Kumar:2026ogn,Kumar:2026dih}, which typically assume the weak mixing limit to define $f_{\rm NL}$ amplitudes and compute the underlying shape functions. The inaccessibility of weak mixing is unlikely to change in the near future: upcoming CMB datasets are expected to improve constraints on (equilateral-type) non-Gaussianity by $(2-5)\times$ \citep{SimonsObservatory:2018koc,CMB-S4:2016ple}, and the sensitivity of large-scale structure probes is currently significantly below that of the CMB \citep{Chudaykin:2025vdh,Cabass:2022epm} (though the situation is beginning to change for complementary series exchange \citep{Green:2023uyz,DAmico:2022gki,Chudaykin:2025vdh,Chaussidon:2024qni,Cabass:2022ymb}).


In the strongly-mixed regime, the prospects for constraining multi-field inflation are somewhat brighter, particularly given the recent influx of theoretical interest \citep{Huenupi:2026abj,Huenupi:2026aqc,Pinol:2026xnl,Wang:2026lff}. As shown in Fig.\,\ref{fig: results-strong-mixing}, the \textit{Planck} bounds on coupling amplitudes are competitive with respect to the theoretical limits from unitarity across a wide range of masses (though are fairly unnatural cf.\,\ref{eq: naturalness}). That said, our analysis gives no evidence for new physics at any mass, mixing, or sound-speed, with a maximum $\Delta\chi^2$ of $5.3$ and an evidence ratio far below unity. We further note that the characteristic collider oscillations are highly suppressed in the strong mixing regime (with exponential damping in $\rho/H$): to fully explore the theoretically exciting regime of $\rho\gtrsim H$, we will likely require futuristic datasets, such as dark ages neutral hydrogen datasets \citep{Munoz:2015eqa,Meerburg:2016zdz,Liu:2022iyy}.

While the conclusions appear bleak for the simplest scalar field models, a number of other inflationary scenarios may produce larger primordial signals. These include scenarios with a chemical potential \citep{Bodas:2020yho,Sou:2021juh,Bodas:2025vpb,Qin:2025xct,Tong:2022cdz,Qin:2025xct}, higher-spin exchange \citep{Bordin:2018pca,Alexander:2019vtb,Tong:2022cdz,Kim:2019wjo} loop corrections \citep{Wang:2021qez,Xianyu:2022jwk,Qin:2023bjk,Qin:2023nhv,Qin:2024gtr,Pimentel:2026kqc,Liu:2026jzn}, curvaton enhancement \citep{Cassem:2026ygh}, models with multiple isocurvatons \citep{McAneny:2019epy,Green:2026yev,Ferreira:2026tyj}, dissipative interactions \citep{Salcedo:2024smn,Salcedo:2026sdn}, strongly-coupled sectors \citep{Jiang:2025mlm,Pimentel:2025rds,Philcox:2026bfa}, tachyonic exchange \citep{McCulloch:2024hiz}, features in the potential \citep[e.g.,][]{Chen:2022vzh,Pinol:2023oux,Werth:2023pfl} and 
non-Bunch-Davies initial states \citep{Meerburg:2009ys,Yin:2023jlv}. These models provide an exciting discovery space that can be realistically probed using the methods developed in this work.

\acknowledgments
{\small
\begingroup
\hypersetup{hidelinks}
\noindent 
We thank Soubhik Kumar, Lucas Pinol, and Denis Werth for insightful discussions. This work was inspired by conversations at the 41st Annual Colloquium of the Institut d'Astrophysique de Paris. OHEP thanks the \href{https://www.flickr.com/photos/198816819@N07/55371776133/in/dateposted-public/}{Chengdu Research Base for stimullating discussions}. The computations in this work were run at facilities supported by the Scientific Computing Core at the Flatiron Institute, a division of the Simons Foundation. 
\endgroup
\vskip 4pt
}

\appendix

\section{Results with an Alternative $\rho$ Prior}\label{app: alternative-prior}
\noindent In the main text, we analyzed the two-field inflationary Lagrangian using a flat prior on the quadratic mixing: $\rho\sim \mathcal{U}(0,\rho_{\rm max}(c_\pi,c_\sigma))$. Here, we present results obtained using the alternative prior $\log_{10}\rho\sim \mathcal{U}(-4,\log_{10}\rho_{\rm max})$. This downweights the $\rho\gg 1$ regime, which leads to significant differences in the parameter posteriors.

\begin{figure}
    \centering
    \includegraphics[width=\linewidth]{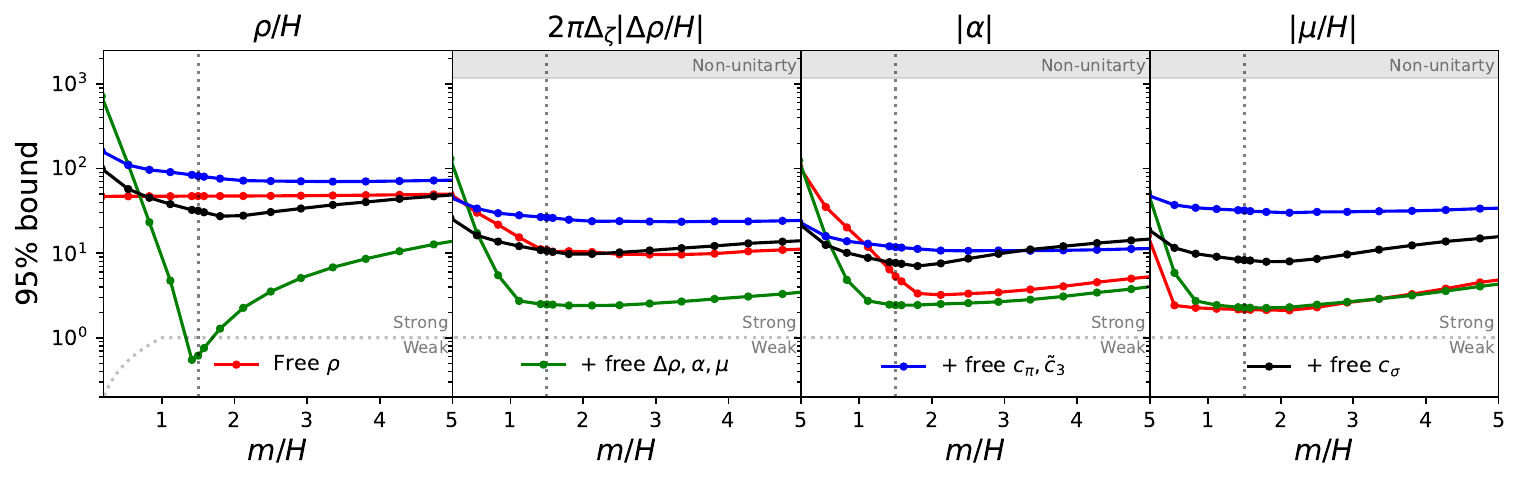}
    \includegraphics[width=\linewidth]{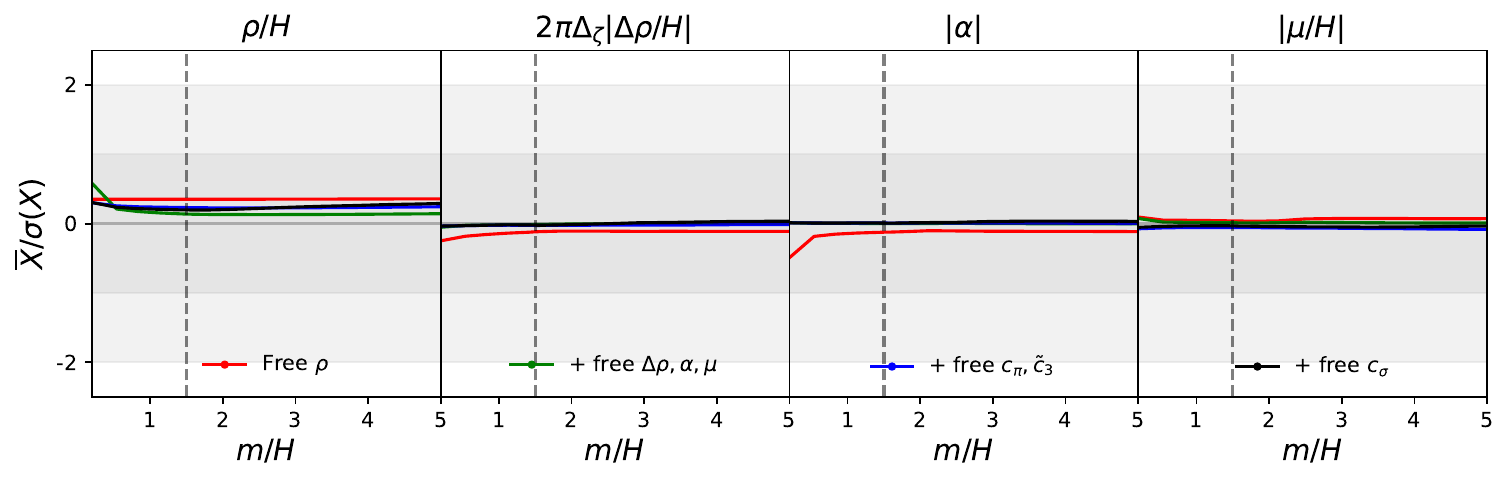}
    \includegraphics[width=0.7\linewidth]{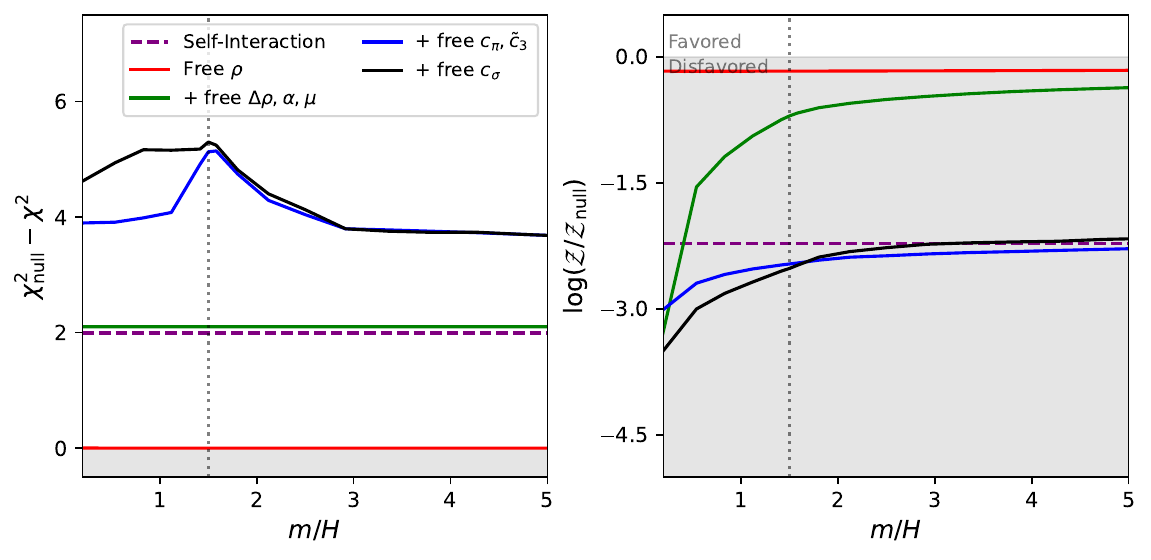}
    \caption{\textbf{Strong Mixing Results (Logarithmic Prior)}: As Fig.\,\ref{fig: results-strong-mixing} but imposing a uniform prior on $\log_{10}\rho$ (instead of on $\rho$). The alternative prior favors small-mixing regimes leading to tighter constraints with increased mass-dependence. The marginalized parameters remain consistent with zero, and we find no evidence for new physics across any of the scenarios considered.} 
    \label{fig: results-strong-mixing-log}
\end{figure}

\begin{figure}
    \centering
    \includegraphics[width=\linewidth]{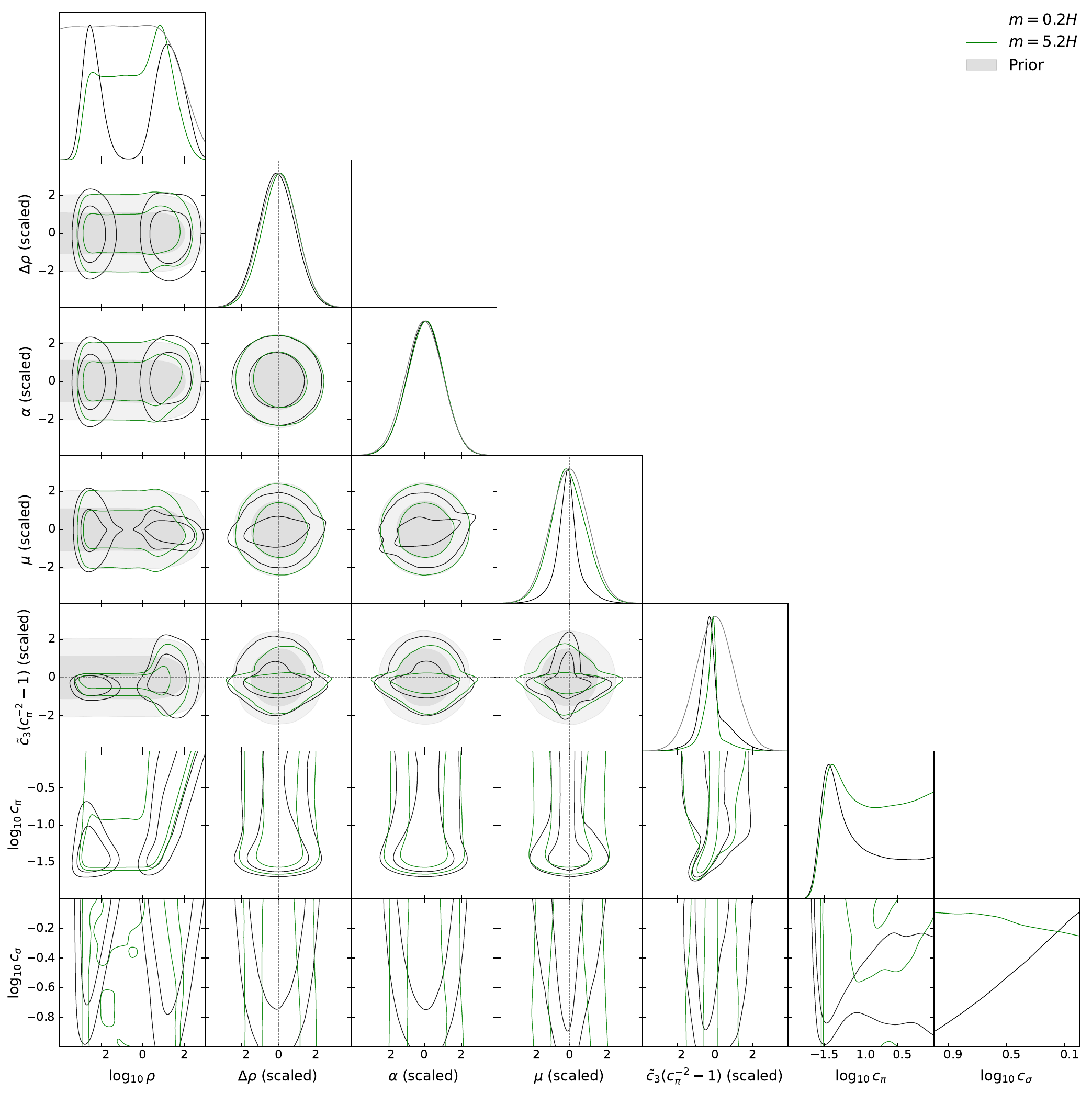}
    \caption{\textbf{Joint Constraints on the Strong Mixing Collider (Logarithmic Prior)}: As Fig.\,\ref{fig: corner-strong-mixing} but with a flat prior on $\log_{10}\rho$. For ease of interpretation, we restrict to two masses: $m=0.2H$ and $m=5.2$. For small masses, we find bimodal behavior in $\log_{10}\rho$, with an excluded region around $\rho\sim 0.1$ (with the prior dominating at small $\rho$), consistent with Fig.\,\ref{fig: results-strong-mixing-fixrho}. At large masses, the constraints on exchange amplitudes are prior-domianted for all $\rho\lesssim H$, leading to the simpler posterior structure.}
    \label{fig: corner-strong-mixing-log}
\end{figure}

The main results are shown in Fig.\,\ref{fig: results-strong-mixing-log} (mirroring Fig.\,\ref{fig: results-strong-mixing}). We find stronger constraints on all parameters, with, for example, the $\rho$-only analysis finding $\rho/H\lesssim 50$. As $m\to 0$, the constraints on $\Delta\rho$, $\alpha$ and $\mu$ weaken significantly, in contrast to the weak mixing scalings of Fig.\,\ref{fig: results-weak-mixing}. This arises since the data place strong constraints on the cubic couplings at low $\rho$ (see Fig.\,\ref{fig: results-strong-mixing-fixrho}), which shifts the posterior volume to large-$\rho$, where the bounds become $m$-independent. This can also be seen from the joint posteriors shown in Fig.\,\ref{fig: corner-strong-mixing-log}: at low $m/H$, we find bimodal behavior in $\log_{10}\rho$, with the constraints transitioning from prior-dominated at $\rho\to 0$ to tightly-constrained at $\rho\approx 0.1H$ and finally to loosely-bound in the strongly-mixed regime ($\rho\gtrsim H$). At large masses, the second contribuition disappears (since the template no longer diverges in the squeezed limit), thus the posterior is unimodal and prior dominated for all $\rho\lesssim H$.

As before, we find no evidence for a non-zero signal with a maximum signal-to-noise of $0.6\sigma$ (for $\rho/H$ at $m=0.2H$). Due to the tight unitarity bounds at low-$\rho$ the Bayesian evidences are less inflated by Occam penalties than for the linear prior: while this leads to larger Bayesian evidences, we still find $\mathcal{Z}<\mathcal{Z}_{\rm null}$, again indicating that the multi-field scenario is disfavored. 



\bibliographystyle{apsrev4-1}
\bibliography{refs}

\end{document}